\documentclass[journal,twoside]{IEEEtran}
\ifCLASSINFOpdf
\else
\fi
\usepackage{multirow}
\usepackage{graphicx}
\usepackage{epstopdf}
\usepackage{stfloats}
\usepackage{subfigure}
\usepackage{microtype}
\usepackage{setspace}
\usepackage{cite}
\usepackage{color}
\usepackage{amsmath}
\usepackage{array}
\usepackage{pifont}
\usepackage{color}
\usepackage{bm}

\begin{document}
%
\title{Machine Learning-Based 3D Channel Modeling for U2V mmWave Communications}
%
%
%

\author{Kai~Mao,~\IEEEmembership{Student Member,~IEEE;} Qiuming~Zhu,~\IEEEmembership{Member,~IEEE;} Maozhong~Song; Hanpeng~Li; Benzhe~Ning; Boyu~Hua,~\IEEEmembership{Member,~IEEE;} Wei~Fan,~\IEEEmembership{Senior Member,~IEEE}

\thanks{This work was supported in part by the NSFC Key Scientific Instrument and Equipment Development Project under Grant, No. 61827801, in part by ISN State Key Laboratory fund, No. ISN22--11, in part by Aeronautical Science Foundation of China, No. 201901052001, and in part by the Fundamental Research Funds for the Central Universities, No. NS2020026 and No. NS2020063. \emph{(Corresponding author: Q.~Zhu, W. Fan.)}}
\thanks{K.~Mao, Q.~Zhu, M.~Song, H.~Li, B.~Ning, and B.~Hua are with The Key Laboratory of Dynamic Cognitive System of Electromagnetic Spectrum Space, College of Electronic and Information Engineering, Nanjing University of Aeronautics and Astronautics, Nanjing 211106, China (e-mail: \{maokai; zhuqiuming; smz108; sz2004013; ningbenzhe; byhua\}@nuaa.edu.cn).}
\thanks{W.~Fan is with the Antenna Propagation and Millimeter-wave Systems (APMS) section, Department of Electronic Systems, Faculty of Engineering and Science, Aalborg University, Aalborg 9220, Denmark (e-mail: wfa@es.aau.dk).}}

\markboth{IEEE Transactions on Antennas \& Propagation, VOL. XX, NO. XX, May 2021}
{Machine Learning-Based 3D Channel Modeling for U2V mmWave Communications}
%



\maketitle

\begin{abstract}
Unmanned aerial vehicle (UAV) millimeter wave (mmWave) technologies can provide flexible link and high data rate for future communication networks. By considering the new features of three-dimensional (3D) scattering space, 3D velocity, 3D antenna array, and especially 3D rotations, a machine learning (ML) integrated UAV-to-Vehicle (U2V) mmWave channel model is proposed. Meanwhile, a ML-based network for channel parameter calculation and generation is developed. The deterministic parameters are calculated based on the simplified geometry information, while the random ones are generated by the back propagation based neural network (BPNN) and generative adversarial network (GAN), where the training data set is obtained from massive ray-tracing (RT) simulations. Moreover, theoretical expressions of channel statistical properties, i.e., power delay profile (PDP), autocorrelation function (ACF), Doppler power spectrum density (DPSD), and cross-correlation function (CCF) are derived and analyzed. Finally, the U2V mmWave channel is generated under a typical urban scenario at 28 GHz. The generated PDP and DPSD show good agreement with RT-based results, which validates the effectiveness of proposed method. Moreover, the impact of 3D rotations, which has rarely been reported in previous works, can be observed in the generated CCF and ACF, which are also consistent with the theoretical and measurement results.
\end{abstract}


%
\IEEEpeerreviewmaketitle

\section{Introduction}
%
%
%
%
\IEEEPARstart{U}{nmanned} aerial vehicle (UAV) are expected to play an important role in the sixth generation (6G) wireless communication networks, and the millimeter wave (mmWave) technology is promising for UAVs to further improve the system capacity and transmission rate \cite{WCX21_SCIS, UllahZ20_TCCN, DingG18_ICM, XZ20_ITJ, ZhongW19_CC}. However, different from terrestrial mmWave communication scenarios, the UAV-to-vehicle (U2V) mmWave scenario has some unique features, i.e., three-dimensional (3D) scattering space, 3D velocity, 3D rotations, and 3D antenna array, which would significantly affect the channel characteristics \cite{Khawaja19_CST, Zhang19_WC, AiB20_TVT}. Therefore, it is of vital importance to design accurate channel models for the design, optimization, and evaluation of U2V mmWave communication systems.
\par UAV related propagation channel models at sub-6GHz band have been reported in \cite{WCX20_Trans, ZhuQ19_IET, GuanK20_ArXiv, ChengX20_ITJ, Khawaja19_Survey, HeR20_Trans}. However, these models did not consider the mmWave propagation properties. For the mmWave band, the channel models under terrestrial scenarios have been well studied \cite{Rappaport21_JSAC, Rappaport19_Globecom, Heath17_WCL}. For example, the authors in \cite{Rappaport21_JSAC, Rappaport19_Globecom} carried out a lot of field measurements in indoor scenarios for mmWave and sub-Terahertz frequencies, i.e., 28, 73, and 140 GHz. A large-scale path loss model and unified indoor statistical channel model were provided based on the measurement data. The authors in \cite{Heath17_WCL} performed several field measurements to examine the impact of body shadowing and movement upon the communication channel at 60 GHz. The channel characteristics, i.e., path loss, shadowing, and small-scale fading were also extracted and analyzed. Recently, the UAV mmWave channel modeling has attracted more and more interests, which mainly includes the field-measurement based model \cite{Geise18_CAMA, Garcia20_Trans, Semkin21_arXiv}, the ray-tracing (RT) based model \cite{Khawaja17_VTC, Khawaja18GSMM, ZhuQ20_EuCAP, ZhuQ20IWCMC} and the geometry-based stochastic model (GBSM) \cite{Michailidis20_TVT, WCX20_ITJ, HeR20_GlobeCom}. For example, air-to-ground channel sounding systems were implemented based on the Octocopter, hexacopter and DJI UAV in \cite{Geise18_CAMA, Garcia20_Trans, Semkin21_arXiv}, and measurement campaigns were performed at 16 GHz, 60Hz, and 28 GHz, respectively. Field-measurement based channel modeling is an accurate method to guarantee realistic channel characteristics for the specific scenario, but the hardware implementation is not only expensive but also extremely complicated due to the limitation of volume and payload weight for UAVs. Moreover, field measurement can only cover limited scenarios and the generality of channel model is restricted. On the other hand, RT is an efficient method for mmWave channel modeling since the mmWave signal has quasi-optical characteristic. For example, the authors in \cite{Khawaja17_VTC, Khawaja18GSMM, ZhuQ20_EuCAP, ZhuQ20IWCMC} proposed several UAV mmWave channel models and analyzed the channel characteristics based on RT method. Note that the accuracy of RT-based models highly depends on the geometric and electromagnetic details of digital map, i.e., shapes, materials, etc. Moreover, it is also time-consuming to run the real-time RT-based model especially for the dynamic U2V scenarios.
\par GBSMs have been widely applied in both sub-6GHz and mmWave channel modeling due to its good balance between complexity and accuracy \cite{ZhuQ18_Trans, ZhuQ19_IET, Michailidis20_TVT, WCX20_ITJ, HeR20_GlobeCom}. However, some UAV mmWave GBSMs were built based on the unrealistic mmWave channel characteristics such as the path number, path power and power-angular spectrum (PAS) following poisson \cite{Michailidis20_TVT}, exponential \cite{Michailidis20_TVT, WCX20_ITJ}, and VonMises \cite{Michailidis20_TVT, HeR20_GlobeCom} distributions, respectively. To overcome this shortcoming, UAV mmWave channel data from RT simulations or field measurements were used to optimize the mmWave channel parameters \cite{Health21_Trans, CaiX21_Trans, Mao20_Sensors}. The authors in \cite{Health21_Trans} proposed a line-of-sight (LoS) probability model and the closed-form expressions are derived based on certain characteristic distributions for UAV mmWave communications. The authors in \cite{CaiX21_Trans} analyzed the delay spread, Doppler frequency spread, and K-factor based on the measurement data. The authors in \cite{Mao20_Sensors} studied the statistical distributions of path power, intra-path delay and angle by RT method. For these research works, based on a given empirical expression, the regression method was adopted to determine parameters by fitting the simulated or measured data. An accurate empirical expression is the key step, but it needs knowledge and expertise of radio propagation. Moreover, it is hard to find a general empirical expression for diverse propagation environments.
\par The aforementioned prior works predict the model parameters from signal processing perspective. Recently, machine learning (ML) has become a new valuable tool by recasting parameter prediction as a learning-based optimization problem \cite{HeRS17_Trans, QiN21_Trans, Aldossari19_WPC, ZhangJH20_CC, WenT21_early, Bharti20_OJAP, Thrane20_Access, Navabi18_ICC, HeR21_Trans,  ZhaoXW20_JSAC, ZhangY19_MAP, William20_arXiv}. For example, the authors in \cite{Navabi18_ICC} applied the neural network to predicted the angle-of-departure (AoD) based on the obtained angle-of-arrival (AoA), while a fast AoA prediction method was proposed in \cite{HeR21_Trans} by using the support vector machine. The authors in \cite{ZhaoXW20_JSAC} modeled the path loss and channel coefficient by using an artificial neural network and integrated it into a GBSM-based channel simulation framework. These research works mainly focused on the conventional terrestrial communications. To our best knowledge, very few ML-based channel parameter estimation or modeling involves UAV mmWave communication scenario \cite{ZhangY19_MAP, William20_arXiv}. The authors in \cite{ZhangY19_MAP} predicted the path loss and delay spread of UAV mmWave channels via a supervised regression algorithm. The generative neural network was utilized in \cite{William20_arXiv} to model the UAV mmWave channel with respect to the line-of-sight (LoS) probability, path loss, and angular distribution, etc. Note that it is hard to observe the effect of unique 3D characteristics on the UAV communication channels since the model was completely driven by data. Moreover, both \cite{ZhangY19_MAP} and \cite{William20_arXiv} did not consider the factor of 3D rotation for the UAV communication scenario. This paper aims to fill this gap. The contributions and novelties of the work are summarized as follows:
\par 1) A ML-based U2V mmWave channel model is proposed, which considers new features of 3D scattering space, 3D velocity, 3D antenna array as well as 3D rotation under realistic U2V communication scenarios. Moreover, the ML networks, i.e., back propagation based neural network (BPNN) and generative adversarial network (GAN), are introduced to model path power and angle to improve the efficiency and accuracy of proposed model, respectively.
\par 2) A hybrid framework of U2V channel generation based on ML method is developed, which consists of deterministic part and random part. The deterministic channel parameters, i.e., path numbers, path delays, and path angles, are calculated by the geometry relationships of user-defined scenario. The random ones, i.e., path powers and ray angle offsets, are predicted by BPNN and GAN, respectively, which are trained by massive RT simulation data. To obtain the training sets of angle offsets, a K-means-based clustering algorithm is developed as well.
\par 3) The statistical properties of proposed U2V mmWave channel model, i.e., power delay profile (PDP), autocorrelation function (ACF), Doppler power spectrum density (DPSD) and cross-correlation function (CCF) are derived and analyzed. Then, the U2V mmWave channel is generated based on the proposed model under a typical urban scenario at 28 GHz. To validate the effectiveness of proposed method, the generated PDP and DPSD are compared with the RT-based ones, and the generated CCF and ACF are compared with the theoretical and measurement ones. The results show that the characteristics of generated UAV mmWave channel are highly consistent with the RT simulated or measured ones. Moreover, the effect of 3D rotation on the CCF is demonstrated and analyzed.
\par The rest paper is organized as follows. In Section II, a ML-based 3D channel model for U2V mmWave communications is proposed. Section III gives the ML-based channel generation scheme and the details of channel parameter generation. Section IV derives the statistical properties of proposed channel model. The simulation and validation of proposed model are given in Section V. Finally, conclusions are drawn in Section VI.

\section{U2V channel model with velocity and attitude variations}
\subsection{Analytical channel model}
Let us consider a typical U2V communication system as shown in Fig. 1, where the UAV is the transmitter denoted by Tx and the vehicle is the receiver denoted by Rx. The UAV is equipped with a $p$-element antenna array and moving with velocity and attitude variations denoted by ${{\bf{v}}^{{\rm{tx}}}}(t)$. The vehicle is equipped with a $q$-element antenna array and moving with velocity variations and constant attitude denoted by ${{\bf{v}}^{{\rm{rx}}}}(t)$. For simplicity of derivation, the Tx and Rx have their own coordinate system with the origins at the central of UAV and vehicle, respectively. It should be mentioned that both the UAV and vehicle are moving in 3D arbitrary trajectories, while the UAV has a time-variant attitude due to the rotation movement. The rotations, i.e., yaw, pitch, and roll, are supported by most of UAVs, but not considered in most of literatures on channel modeling \cite{Michailidis20_TVT, WCX20_ITJ, HeR20_GlobeCom}.

\begin{figure}[!tb]
	\centering
	\includegraphics[width=88mm]{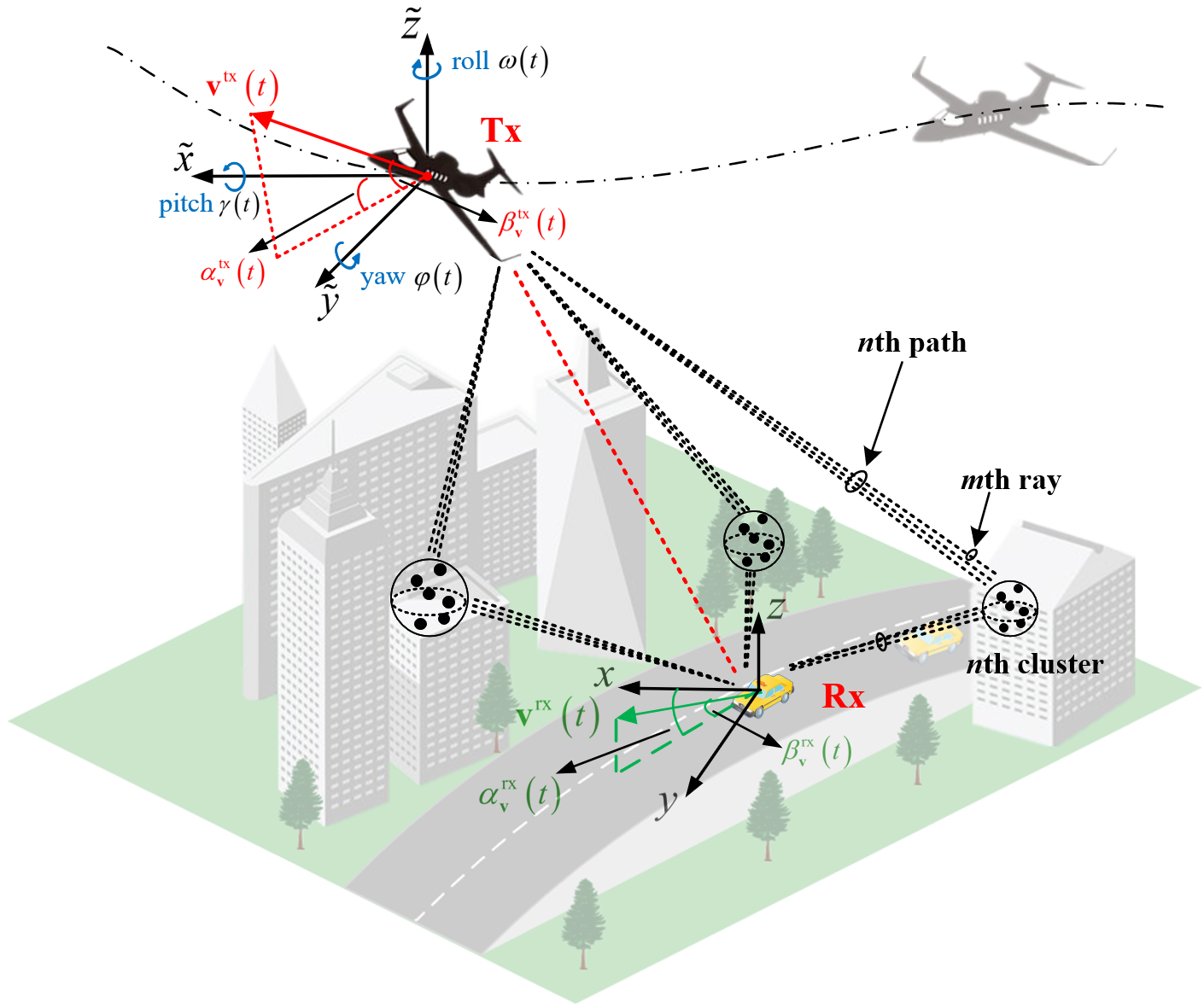}
	\caption{Typical U2V mmWave communication system.}
    \label{fig:1}
\end{figure}

\par Based on the theory of radio propagation, the U2V channel impulse response (CIR) between the $p$th Tx antenna and the $q$th Rx antenna can be modelled by the summation of several paths as
\begin{equation}
{h_{qp}}\left( {t,\tau } \right){\rm{ = }}\sum\limits_{n = 1}^{N(t)} {\sqrt {{P_n}\left( t \right)} {{\widetilde h}_n}\left( t \right)\delta \left( {\tau  - {\tau _n}\left( t \right)} \right)}
\label{0}
\end{equation}
\noindent where $N\left( t \right)$ is the number of valid path,  ${P_n}\left( t \right)$ and ${\tau _n}\left( t \right)$ are the power gain and delay of each path (simplified as path power and path delay later), respectively. Moreover, we ignore the antenna index $q$, $p$ of the other parameters for simplicity. In the realistic environment, the valid propagation path could be line-of-sight (LoS) or non-line-of-sight (NLoS), i.e., single-, double- and multiple-bounce. However, mmWave channels have been found to be sparse and specular \cite{FanW16_JWCN, FanW19_TVT, HeR21_WC}, and this phenomenon is more pronounced in UAV mmWave channels \cite{Mao20_Sensors}. Thus, Only the LoS path and single-bounce NLoS paths are considered in this paper. In (1), ${\widetilde h_n}\left( t \right)$ represents the normalized complex coefficient (also known as small scale channel fading) of $n$th path where the path is defined as the combination of clustered propagation rays with similar delays and powers but different angles, e.g., azimuth angle of departure (AAoD), elevation angle of departure (EAoD), azimuth angle of arrival (AAoA), and elevation angle of arrival (EAoA). Here, the normalized complex coefficient is expressed as
\begin{equation}
\begin{aligned}
{\widetilde h_n}\left( t \right) &= \sqrt {\frac{1}{M}} \sum\limits_{m = 1}^M {{e^{{\rm{j}}\psi _{n,m}^{}\left( t \right)}}} \\
 &= \sqrt {\frac{1}{M}} \sum\limits_{m = 1}^M {{e^{{\rm{j}}\left( {\psi _{n,m}^{\rm{I}}{\rm{ + }}\psi _{n,m}^{\rm{D}}\left( t \right) + \psi _{n,m}^{\rm{R}}\left( t \right)} \right)}}}
\end{aligned}
\label{1}
\end{equation}
\noindent where $M$ is the ray number within the $n$th path and it is usually 1 for the LoS case and a large value for the NLoS case, and $\psi _{n,m}^{}\left( t \right)$ is the time-variant phase of each ray. For taking the effects of velocity and attitude on the phase into account, the time-variant phase is modeled by three part, i.e., $\psi _{n,m}^{\rm{I}}$, $\psi _{n,m}^{\rm{D}}\left( t \right)$, and $\psi _{n,m}^{\rm{R}}\left( t \right)$, which are the initial phase and the phases caused by movement and rotation, respectively. Note that the power gain ${P_n}$  (also known as the large scale channel fading) is extracted from each path and the total power of each path is normalized as one. Thus, the amplitude of $M$ rays within each path can be averaged as $\sqrt{1/M}$.

\subsection{ML-based channel model}
In previous works, the channel parameters in (1) were obtained in a deterministic way based on the specific digital map \cite{Khawaja18GSMM, ZhuQ20_EuCAP, ZhuQ20IWCMC} or in a stochastic way by using the empirical results \cite{3GPP18, ZhuQ19_3GPP, WINNER+, 5GCM}. Note that it is time-consuming or sometimes even impractical to generate the U2V channels with the deterministic method since the propagation scenario changes rapidly due to the velocity and attitude variations. On the other hand, there are very limited empirical results for U2V mmWave channels due to the complexity and high cost associated with channel sounding campaigns. Thus, it is quite difficult to describe the stochastic channel parameters in an accurate way.
\par To improve the accuracy and efficiency of U2V channel modeling and generation, a ML-based framework is introduced to upgrade the above analytical channel model. Our basic idea is to build a new ML-based channel model, where the channel parameters are described and obtained in real-time by several ML-based networks. For example, the BPNN and GAN are applied to generate the path power and ray angle, respectively. All networks can be trained in advance by RT simulation data or field measurements. The proposed ML-based U2V mmWave channel model can be expressed as
\normalsize
\begin{equation}
\begin{array}{c}
{h_{qp}}\left( {t,\tau ;{\bf{w}},{\bf{b}},{\bm{\sigma }};{{\bm{\theta }}_g}} \right){\rm{ = }}\sum\limits_{n = 0}^{N\left( t \right)} {\{ \sqrt {{P_n}\left( {t;{\tau _n},{\bf{w}},{\bf{b}},{\bm{\sigma }}} \right)} } \\
 \ \ \ \ \ \ \ \ \ \ \ \ \ \ \ \ \ \ \ \ \ \ \ \ \ \ \cdot {\widetilde h_n}\left( {t;{{\bm{\theta }}_g}} \right)\delta \left( {\tau  - {\tau _n}\left( t \right)} \right)\}
\end{array}
\label{2}
\end{equation}
\noindent where $\bf{w}$, $\bf{b}$ and $\bm{\sigma}$ are the weight matrix, bias matrix, and activation matrix of BPNN, respectively, and ${{\bm{\theta}}_g}$ is the equivalent parameter matrix of generator network in GAN. Similarly, the channel coefficient can be rewritten as
\begin{equation}
{\widetilde h_n}\left( {t;{{\bm{\theta }}_g}} \right) = \sqrt {\frac{1}{M}} \sum\limits_{m = 1}^M {{e^{{\rm{j}}\left( {\psi _{n,m}^{\rm{I}}\left( {{t_0}} \right){\rm{ + }}\psi _{n,m}^{\rm{D}}\left( {t;{{\bm{\theta }}_g}} \right) + \psi _{n,m}^{\rm{R}}\left( {t;{{\bm{\theta }}_g}} \right)} \right)}}}
\label{3}
\end{equation}
where the initial phase at time instant ${t_0}$ and phase caused by Doppler frequency can be calculated respectively as
\begin{equation}
\psi _{n,m}^{\rm{I}}\left( {{t_0}} \right) = \int_0^{{t_0}} {2{\rm{\pi}}{f_{n,m}}\left( {t^{'}} \right)} {{\rm{d}}t^{'}} + {\psi^{'}}_{n,m}^{\rm{I}}
\label{4}
\end{equation}
\begin{equation}
\psi _{n,m}^{\rm{D}}\left( {t;{{\bm{\theta }}_g}} \right) = \int_{{t_0}}^t {2{\rm{\pi }}{f_{n,m}}\left( {t';{{\bm{\theta }}_g}} \right)} {\rm{d}}t^{'}
\label{5}
\end{equation}
\noindent where ${\psi^{'}}_{n,m}^{\rm{I}}$ represents the random initial phase distributed uniformly over $[0,2\pi )$, ${f_{n,m}}(t)$ is the Doppler frequency and can be further calculated as
\begin{equation}
\begin{aligned}
{f_{n,m}}(t;{{\bm{\theta }}_g}) &= \frac{{{f_0}}}{c}{\bf{r}}_{n,m}^{{\rm{tx}}}\left( {t;{{\bm{\theta }}_g}} \right){\left( {{{\bf{v}}^{{\rm{tx}}}}\left( t \right)} \right)^{\rm{T}}}\\
&\ \ {\rm{ + }}\frac{{{f_0}}}{c}{\bf{r}}_{n,m}^{{\rm{rx}}}\left( {t;{{\bm{\theta }}_g}} \right){\left( {{{\bf{v}}^{{\rm{rx}}}}\left( t \right)} \right)^{\rm{T}}}
\end{aligned}
\label{6}
\end{equation}
\noindent where ${f_0}$ is the carrier frequency, $c$ is the light speed, and ${{\bf{v}}^{{\rm{tx}}}}\left( t \right)$, ${{\bf{v}}^{{\rm{rx}}}}\left( t \right)$ are the velocity vector of Tx and Rx, respectively. In (7), ${\bf{r}}_{n,m}^{{\rm{tx/rx}}}(t)$ is the spherical unit vectors and can be written as
\begin{equation}
{\bf{r}}_{n,m}^{{\rm{tx/rx}}}(t;{{\bm{\theta }}_g}) = {\left[ {\begin{array}{*{20}{c}}
{\cos \beta _{n,m}^{{\rm{tx/rx}}}(t;{{\bm{\theta }}_g})\cos\alpha _{n,m}^{{\rm{tx/rx}}}(t;{{\bm{\theta }}_g})}\\
{\cos \beta _{n,m}^{{\rm{tx/rx}}}(t;{{\bm{\theta }}_g})\sin \alpha _{n,m}^{{\rm{tx/rx}}}(t;{{\bm{\theta }}_g})}\\
{\sin \beta _{n,m}^{{\rm{tx/rx}}}(t;{{\bm{\theta }}_g})}
\end{array}} \right]^{\rm{T}}}
\label{7}
\end{equation}
\noindent where $\alpha _{n,m}^{{\rm{tx}}}(t;{{\bm{\theta }}_g})$, $\beta _{n,m}^{{\rm{tx}}}(t;{{\bm{\theta }}_g})$ are the AAoD and EAoD of propagation ray, respectively, and $\alpha _{n,m}^{{\rm{rx}}}(t;{{\bm{\theta }}_g})$, $\beta _{n,m}^{{\rm{rx}}}(t;{{\bm{\theta }}_g})$ are AAoA and EAoA of propagation ray, respectively. Moreover, the phase caused by rotation can be calculated by
\begin{equation}
\begin{aligned}
\!\!\!\! \psi _{n,m}^{\rm{R}}(t;{{\bm{\theta }}_g}) &= \frac{{2\pi {f_0}}}{c}\left( {{\bf{r}}_{n,m}^{{\rm{rx}}}(t;{{\bm{\theta }}_g}) \cdot {\bf{R}}_{\bf{v}}^{{\rm{rx}}}(t) \cdot {{\bf{d}}^{{\rm{rx}}}}(t)} \right)\\
 &\!\!\!\!\!+ \frac{{2\pi {f_0}}}{c}\left( {{\bf{r}}_{n,m}^{{\rm{tx}}}(t;{{\bm{\theta }}_g}) \cdot {\bf{R}}_{\bf{v}}^{{\rm{tx}}}(t) \cdot {{\bf{R}}^{\rm{P}}}(t) \cdot {{\bf{d}}^{{\rm{tx}}}}(t)} \right)
\label{8}
\end{aligned}
\end{equation}

\begin{figure*}[!b]
\normalsize
\begin{equation}
{\bf{R}}_{\bf{v}}^{{\rm{tx/rx}}}\left( t \right) = \left[ {\begin{array}{*{20}{c}}
{\cos \alpha _{\rm{v}}^{{\rm{tx/rx}}}\left( t \right)\cos \beta _{\rm{v}}^{{\rm{tx/rx}}}\left( t \right)}&{ - \sin \alpha _{\rm{v}}^{{\rm{tx/rx}}}\left( t \right)}&{ - \cos \alpha _{\rm{v}}^{{\rm{tx/rx}}}\left( t \right)\sin \beta _{\rm{v}}^{{\rm{tx/rx}}}\left( t \right)}\\
{\sin \alpha _{\rm{v}}^{{\rm{tx/rx}}}\left( t \right)\cos \beta _{\rm{v}}^{{\rm{tx/rx}}}\left( t \right)}&{\cos \alpha _{\rm{v}}^{{\rm{tx/rx}}}\left( t \right)}&{ - \sin \alpha _{\rm{v}}^{{\rm{tx/rx}}}\left( t \right)\sin \beta _{\rm{v}}^{{\rm{tx/rx}}}\left( t \right)}\\
{\sin \beta _{\rm{v}}^{{\rm{tx/rx}}}\left( t \right)}&0&{\cos \beta _{\rm{v}}^{{\rm{tx/rx}}}\left( t \right)}
\end{array}} \right]
\label{9}
\end{equation}
\end{figure*}
\begin{figure*}[!b]
\normalsize
\begin{equation}
{{\bf{R}}^{\rm{P}}}\left( t \right) = \left[ {\begin{array}{*{20}{c}}
{\cos \left( \omega  \right)\cos \left( \varphi  \right)}&{\cos \left( \omega  \right)\sin \left( \varphi  \right)\sin \left( \gamma  \right) - \sin \left( \omega  \right)\cos \left( \gamma  \right)}&{\cos \left( \omega  \right)\sin \left( \varphi  \right)\cos \left( \gamma  \right) + \sin \left( \omega  \right)\sin \left( \gamma  \right)}\\
{\sin \left( \omega  \right)\cos \left( \varphi  \right)}&{\sin \left( \omega  \right)\sin \left( \varphi  \right)\sin \left( \gamma  \right) + \cos \left( \omega  \right)\cos \left( \gamma  \right)}&{\sin \left( \omega  \right)\sin \left( \varphi  \right)\cos \left( \gamma  \right) - \cos \left( \omega  \right)\sin \left( \gamma  \right)}\\
{ - \sin \left( \varphi  \right)}&{\cos \left( \varphi  \right)\sin \left( \gamma  \right)}&{\cos \left( \varphi  \right)\cos \left( \gamma  \right)}
\end{array}} \right]
\label{10}
\end{equation}
\end{figure*}

\noindent where ${{\bf{d}}^{{\rm{tx/rx}}}}(t)$ is the location vector of Tx (or Rx) antenna.
\noindent It should be noted that the matrix ${\bf{R}}_{\bf{v}}^{{\rm{tx/rx}}}(t)$ is introduced to include the factor of velocity variations at both terminals, while the matrix ${{\bf{R}}^{\rm{P}}}\left( t \right)$ is used to describe the effect of attitude variation that the UAV may experience. These two rotation matrixes can be expressed respectively as (10) and (11),
\noindent where $\alpha _{\rm{v}}^{{\rm{tx/rx}}}$ and $\beta _{\rm{v}}^{{\rm{tx/rx}}}$ are the AAoD (or AAoA) and EAoD (or EAoA) of movement direction, respectively, $\varphi$, $\gamma $, and $\omega$ are angle of yaw, pitch, and roll, respectively. Note that the 3D attitude variation due to rotation movement has an impact on the phase of multipath, which is an important part of the normalized channel coefficient, i.e., small-scale fading according to the Eq. (4). The main channel parameters are listed in Table I.

\begin{table}[!t]
\centering
\caption{Parameter definitions in the proposed model}
\centering
\renewcommand\arraystretch{1.3}
\begin{tabular}{p{2.5cm}<{\centering}p{5.5cm}<{\centering}}
\hline
Parameters & Definition \\ \hline
$N,M$ & path and ray numbers, respectively \\
${\tau_n}(t)$ & path delay \\
${P_n}\left( {t;{\tau _n},{\bf{w}},{\bf{b}},{\bm{\sigma }}} \right)$ & BPNN for path power \\
${\bf{w}},{\bf{b}},{\bm{\sigma }}$ & weight, bias, and activation matrix in BPNN, respectively  \\
${\tilde h_n}\left( {t;{{\bm{\theta }}_g}} \right)$ & GAN for normalized channel coefficient \\
${{\bm{\theta }}_g}$ & equivalent parameter matrix of generator network in GAN \\
$\psi _{n,m}^{\rm{I}}\left( {{t_0}} \right)$ & initial random phase \\
$\psi _{n,m}^{\rm{D}}\left( {t;{{\bm{\theta }}_g}} \right)$ & time-variant phase caused by movement  \\
$\psi _{n,m}^{\rm{R}}(t;{{\bm{\theta }}_g})$ & time-variant phase caused by rotation \\
${\bf{R}}_{\bf{v}}^{{\rm{tx/rx}}}(t)$ & matrix caused by velocity variation\\
${{\bf{R}}^{\rm{P}}}\left( t \right)$ & matrix caused by attitude variation \\
\hline
\end{tabular}
\label{table I}
\end{table}

\section{ML-based network of channel parameters and generation}
\subsection{ML-based channel generation framework}
The channel parameters, i.e., path number, path delay, path power, and ray angle, are essential to generate the U2V channel. Based on the geometry relationship and ML theory, the framework of channel generation is shown in Fig. 2. Firstly, according to the geometry information of user-defined scenario, e.g., initial location, velocity, UAV attitude, etc., the path delay and angle can be calculated. Here, we use the intersection algorithm \cite{Moller97_GT} to calculate the reflection and refraction points according to the geometry relationship between transceivers and scatterers, which can be used to determine the valid path number, delays and angles. Secondly, path delays are used to drive a trained BPNN to get the corresponding path powers. The random angle offsets are produced by a trained GAN. By summing up path angles and angle offsets, the channel coefficients can be further obtained. Finally, the U2V mmWave channel can be generated. Note that the proposed generation method can get good trade-off between accuracy and efficiency. For the inter-path parameters, they are mainly determined by the locations of transceivers and scattering points. Since the locations can be obtained from the given digital map, the geometric information can be applied to calculate the inter-path parameters accurately. For the intra-path parameters, they are mainly related with the diffuse reflection due to the rough surface. It is hard to reconstruct the realistic rough surfaces such as walls, cars, etc. Moreover, RT method consumes too much time if all scattering rays are tracked and calculated. Therefore, the scattering rays and intra-path parameters are generated by ML methods in a stochastic way in this paper.

\begin{figure}[!t]
	\centering
	\includegraphics[width=88mm]{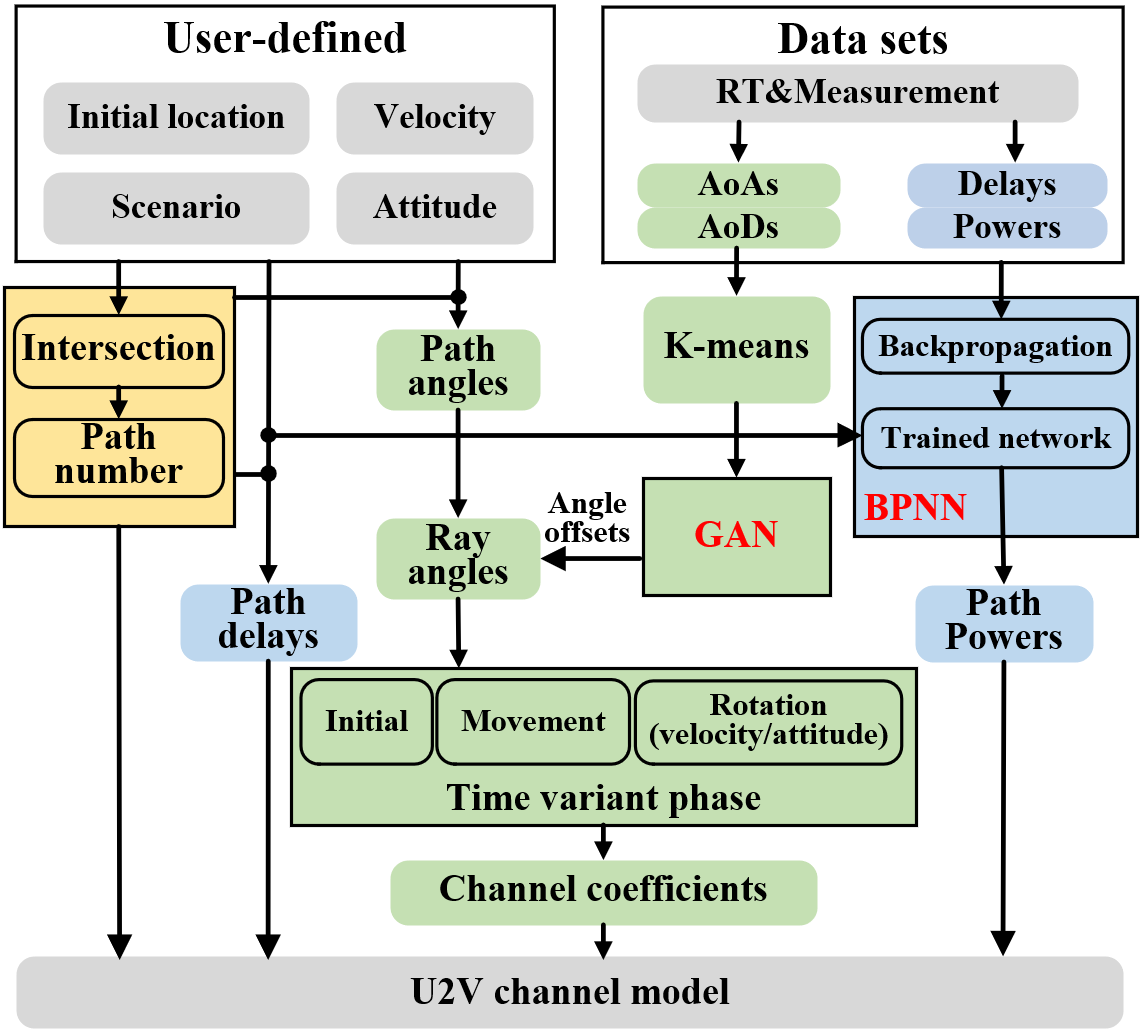}
	\caption{ML-based U2V mmWave channel generation.}
    \label{fig:2}
\end{figure}

\begin{figure*}[!b]
	\centering
	\includegraphics[width=120mm]{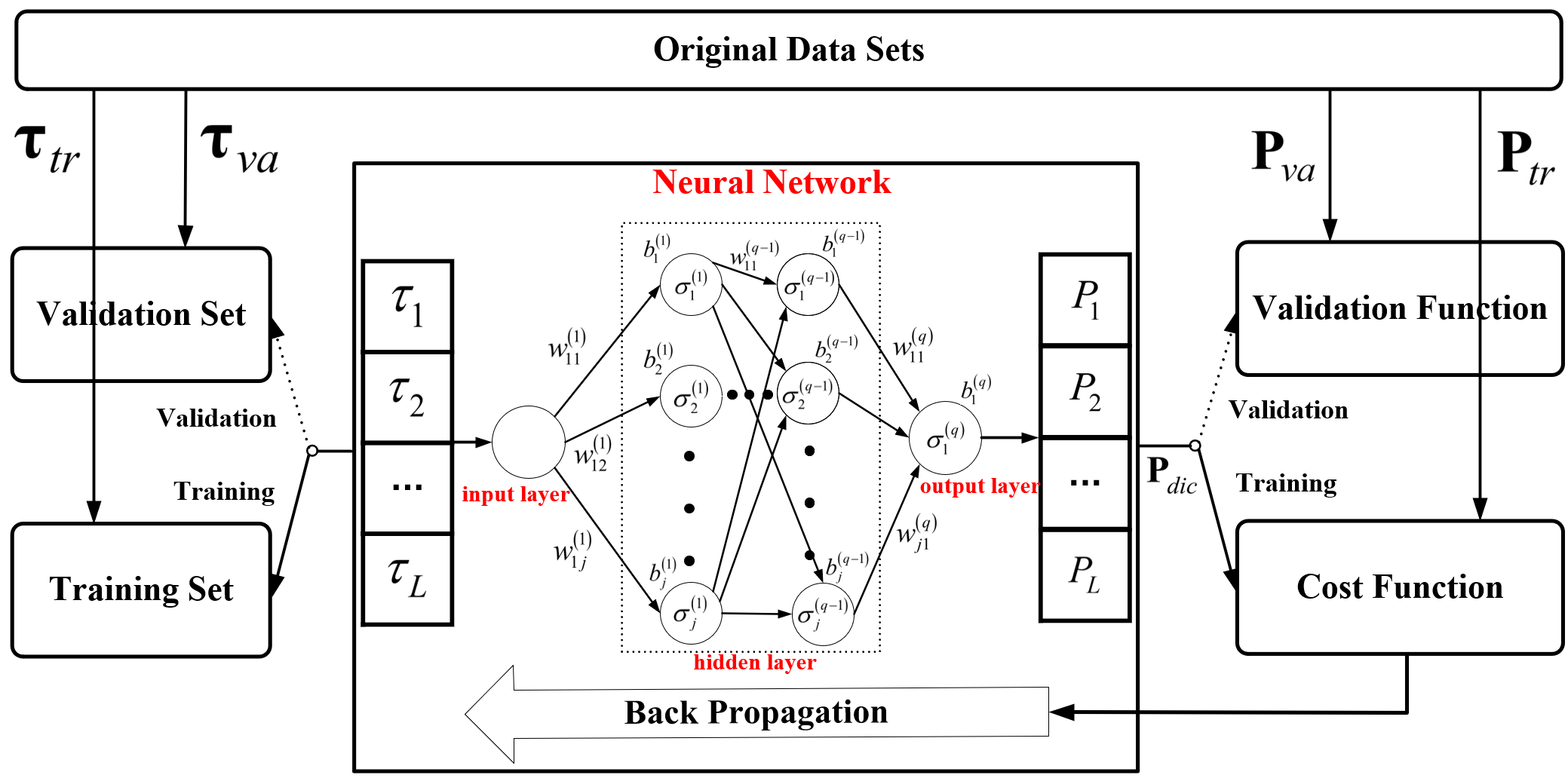}
	\caption{BPNN-based framework of power generation.}
    \label{fig:3}
\end{figure*}

\par On the other hand, ML networks inside the generation framework need to be preprocessed and trained in advance. The training data set can be obtained from either RT simulations or field measurements. For the path delay and power, BPNN is applied to find the inner relationship between them and generate corresponding power according to the input delay in real-time. For the random ray angle offsets, GAN is applied as a generative network with an underlying distribution and generate the variable with the same characteristics as the training data. Since the angle offsets cannot be obtained from the training data set directly, we use the K-means algorithm to cluster the channel data and then get the training data of AoAs and AoDs. It should be mentioned that path power is dependent on the locations of transceivers and specific communication scenario. Therefore, the trained BPNN ensures both scenario-consistency and efficiency to generate the path power according to the geometry-based path delay. Meanwhile, the ray angle offsets are random due to the diffuse reflection on rough surface. Therefore, it is more practical and efficient to use the GAN to find the underlying statistical distribution and then generate random offsets.

\subsection{BPNN-based path delay and power generation}
Let us set the initial locations of Tx (or Rx) as ${{\bf{L}}_{{\rm{tx/rx}}}}\left( {{t_0}} \right)$ and both Tx and Rx are moving along 3D trajectories with velocity variation. The instant location vector of Tx (or Rx) can be calculated by
\begin{equation}
{{\bf{L}}_{{\rm{tx/rx}}}}\left( t \right) = {{\bf{L}}_{{\rm{tx/rx}}}}\left( {{t_0}} \right) + \int_{{t_0}}^t {{{\bf{v}}_{{\rm{tx/rx}}}}\left( {t'} \right){\rm{d}}t'}
\label{11}
\end{equation}

\noindent where ${\left(  \cdot  \right)_{{\rm{tx/rx}}}}$ represents the index of Tx (or Rx). The distance vectors can be further defined respectively as
\begin{equation}
{\bf{D}}_{{\rm{tx/rx }}}^{{\rm{LoS}}}\left( t \right) = {\bf{L}}_{{\rm{tx/rx}}}^{}\left( t \right) - {\bf{L}}_{{\rm{rx/tx}}}^{}\left( t \right)
\label{12}
\end{equation}
\begin{equation}
{\bf{D}}_{{\rm{tx/rx}}}^{{\rm{NLoS, }}n}\left( t \right) = {\bf{L}}_{{\rm{tx/rx}}}^{}\left( t \right) - {\bf{L}}_n^{}\left( t \right)
\label{13}
\end{equation}

\noindent where ${\bf{D}}_{{\rm{tx/rx }}}^{{\rm{LoS}}}\left( t \right)$ represents the travel vector from Tx (or Rx) to Rx (or Tx), ${\bf{D}}_{{\rm{tx/rx}}}^{{\rm{NLoS, }}n}\left( t \right)$ represents the travel vector from Tx (or Rx) to the $n$th cluster, and ${{\bf{L}}_n}\left( t \right)$ denotes the centroid location of $n$th cluster. Thus, the path delays of LoS path and NLoS path can be calculated respectively by
\begin{equation}
\tau _{}^{{\rm{LoS}}}(t) = \frac{{\left\| {{\bf{D}}_{{\rm{tx/rx}}}^{{\rm{LoS}}}(t)} \right\|}}{c}
\label{14}
\end{equation}

\begin{equation}
\tau _n^{{\rm{NLoS}}}(t) = \frac{{\left\| {{\bf{D}}_{{\rm{tx}}}^{{\rm{NLoS, }}n}\left( t \right)} \right\| + \left\| {{\bf{D}}_{{\rm{rx}}}^{{\rm{NLoS, }}n}\left( t \right)} \right\|}}{c}
\label{15}
\end{equation}

\noindent where $\left\|  \cdot  \right\|$ represents the absolute value and $c$ is the speed of light. Note that the geometry information such as the initial locations can be obtained from the user-defined scenario such as the digital map or reconstructed map \cite{ZhuQ20IWCMC, FanW16_EuCAP}.
\par Different from the path delay, the path power cannot be calculated directly based on the geometry information.
As an alternative method, RT is a useful tool for predicting path power. However, RT simulations are site-specific and only valid for the scenarios of interest. Meanwhile, it is time-consuming to obtain massive channel data using RT method in highly dynamic scenario. To address this issue, ML algorithm can be utilized to effectively generate massive channel data by finding the inner relationship of limited RT simulation data. Moreover, some empirical expression for power prediction with respect to delay were studied based on the measurement results and used to calculate the path power \cite{ZhuQ19_3GPP, WINNER+}, but they were only suitable for the land mobile communication under several scenarios. In this paper, we combine this idea and ML algorithm by building a BPNN-based neural network to generate path power with input delay and training it based on the RT simulation data.
\par The BPNN-based generation framework of path power is shown in Fig. 3. Firstly, the original data sets $\left\{ {{\bf{P}},{\bm{\tau }}} \right\}$ from RT simulations or field measurements are divided into two parts by the proportion of 7:3 randomly, i.e., training set $\left\{ {{{\bf{P}}_{tr}},{{\bm{\tau }}_{tr}}} \right\}$ and validation set $\left\{ {{{\bf{P}}_{val}},{{\bm{\tau }}_{val}}} \right\}$. The training set is used to train the parameters of the network neurons and the validation set is used to validate whether the trained network is overfitting or underfitting. It should be noted that there is no over-lap between the training data and validation data, i.e., $\left\{ {{{\bf{P}}_{tr}},{{\bm{\tau }}_{tr}}} \right\} \cap \left\{ {{{\bf{P}}_{val}},{{\bm{\tau }}_{val}}} \right\} = \emptyset $. As a main part of BPNN, the neural network normally consists of input layer, hidden layer, and output layer. Take the neural network with one input layer, one hidden layer with $J$ neurons and one output layer as an example, the generated power can be expressed as

\begin{equation}
{P_{dic{\rm{, }}l}}\left( {\tau ;{\bf{w}},{\bf{b}},{\bm{\sigma }}} \right) = \sigma _1^{\left( 2 \right)}\left[ {\sum\limits_{j = 1}^J {w_{j1}^{\left( 2 \right)}\sigma _j^{\left( 1 \right)}\left( {w_{1j}^{\left( 1 \right)}{\tau _l} + b_j^{\left( 1 \right)}} \right) + b_1^{\left( 2 \right)}} } \right]
\label{16}
\end{equation}

\noindent where $w_{ij}^{\left( q \right)} \in {\bf{w}}$ is the connection weight between the $j$th neuron in the $q$th layer and the $i$th neuron in former layer, and $b_j^{\left( q \right)} \in {\bf{b}}$ is the bias of $j$th neuron in the $q$th layer, $\sigma _j^{\left( q \right)}\left(  \cdot  \right) \in {\bm{\sigma }}$ is the activation function of $j$th neuron in the $q$th layer. Note that the activation functions are introduced into our proposed network for solving arbitrary nonlinear problems. The sigmoid function and leaky rectified linear unit (LeakyReLU) are two normal activation functions and they can be expressed respectively as
\begin{equation}
{\sigma _{sig}}\left( x \right) = \frac{1}{{1 + {{\rm{e}}^{ - x}}}}
\label{17}
\end{equation}
\begin{equation}
{\sigma _{Leak}}\left( x \right) = \max \left( {0,x} \right) + {a^{Leak}}\min \left( {0,x} \right)
\label{18}
\end{equation}
\noindent where ${a^{Leak}}$ is a nonzero slope.

\par For the proposed BPNN network, the cost function is essential. Since the fundamental target is to adjust the network parameters to minimize the value of cost function, the mean-square error (MSE) based cost function considering L2 regularization is designed and it can be expressed as
\begin{equation}
\begin{aligned}
C\left( {{\bf{w}},{\bf{b}},{\bf{\sigma }}} \right) &= \frac{1}{L}\sum\limits_{l = 1}^L {{{\left( {{P_{dic{\rm{, }}l}}\left( {{\tau _{tr{\rm{, }}l}}} \right) - {P_{tr,l}}} \right)}^2}} \\
& \ \ \ + \frac{1}{2}\lambda \sum\limits_{k = 1}^K {{{\left\| {{\bf{w}}_{}^{\left( k \right)}} \right\|}^2}}
\end{aligned}
\label{19}
\end{equation}
\noindent with ${\tau _{tr{\rm{, }}l}} \in {{\bm{\tau }}_{tr}}$, ${P_{tr,l}} \in {{\bf{P}}_{tr}}$, and $\lambda $ is the regularization factor. At the beginning of BPNN training, the weight matrix and bias matrix are set as the initial values. Then, the back propagation algorithm is used to update the network. After being trained, the BPNN is likely to suffer from being under overfitting or underfitting problems. Therefore, the validation function is used to evaluate the performance of network based on the validation data set and can be expressed as

\begin{equation}
{R_{val}} = \sqrt {\frac{1}{L}\sum\limits_{l = 1}^L {{{\left( {{P_{dic{\rm{, }}l}}\left( {{\tau _{val,{\rm{ }}l}}} \right) - {P_{val{\rm{, }}l}}} \right)}^2}} }
\label{20}
\end{equation}
\noindent with ${\tau _{val{\rm{, }}l}} \in {{\bm{\tau }}_{val}}$, ${P_{val,l}} \in {{\bf{P}}_{val}}$.
\subsection{GAN-based angle generation}
In the proposed U2V channel model, each path consists of finite propagation rays due to limited angular resolution in the application. In practice, angle information for each ray is also not possible in measurement due to limited spatial angle resolution offered by antenna system \cite{ZhangXF17_ICL, ZhangXF10_ICL}. In RT simulation, this information, though available, might be not accurate as well, due to lack of exact geometric and electromagnetic description of the environment. In this paper, we only use the given information to deterministically calculate the path angles that can be viewed as the mean angle of clustered rays. By using the signal travel vectors, the mean AAoD, AAoA, EAoD, and EAoA of path angles can be calculated respectively as (22) and
\begin{figure*}[!t]
\normalsize
\begin{equation}
\alpha _n^{{\rm{tx/rx}}}(t) = \left\{ \begin{array}{l}
\arccos (\frac{{\left\| {{\bf{D}}_{{\rm{tx/rx}}}^{{\rm{LoS/NLoS, }}x}(t)} \right\|}}{{\sqrt {{{\left\| {{\bf{D}}_{{\rm{tx/rx }}}^{{\rm{LoS/NLoS, }}x}(t)} \right\|}^2} + {{\left\| {{\bf{D}}_{{\rm{tx/rx }}}^{{\rm{LoS/NLoS, }}y}(t)} \right\|}^2}} }}),{\bf{D}}_{{\rm{tx/rx}}}^{{\rm{LoS/NLoS, }}x}(t) \ge 0\\
\pi  - \arccos (\frac{{\left\| {{\bf{D}}_{{\rm{tx/rx }}}^{{\rm{LoS/NLoS, }}x}(t)} \right\|}}{{\sqrt {{{\left\| {{\bf{D}}_{{\rm{tx/rx}}}^{{\rm{LoS/NLoS, }}x}(t)} \right\|}^2} + {{\left\| {{\bf{D}}_{{\rm{tx/rx}}}^{{\rm{LoS/NLoS, }}y}(t)} \right\|}^2}} }}),{\bf{D}}_{{\rm{tx/rx}}}^{{\rm{LoS/NLoS, }}x}(t) < 0
\end{array} \right.
\label{21}
\end{equation}
\end{figure*}

\begin{equation}
\beta _n^{{\rm{tx/rx}}}(t) = \arcsin (\frac{{\left\| {{\bf{D}}_{{\rm{tx/rx }}}^{{\rm{LoS/NLoS, }}z}(t)} \right\|}}{{\left\| {{\bf{D}}_{{\rm{tx/rx }}}^{{\rm{LoS/NLoS}}}(t)} \right\|}})
\label{22}
\end{equation}

\noindent where ${\left(  \cdot  \right)^x}$, ${\left(  \cdot  \right)^y}$, and ${\left(  \cdot  \right)^z}$ represent the $x$, $y$, and $z$ component of the distance vectors, respectively.

\begin{figure*}[!b]
	\centering
	\includegraphics[width=120mm]{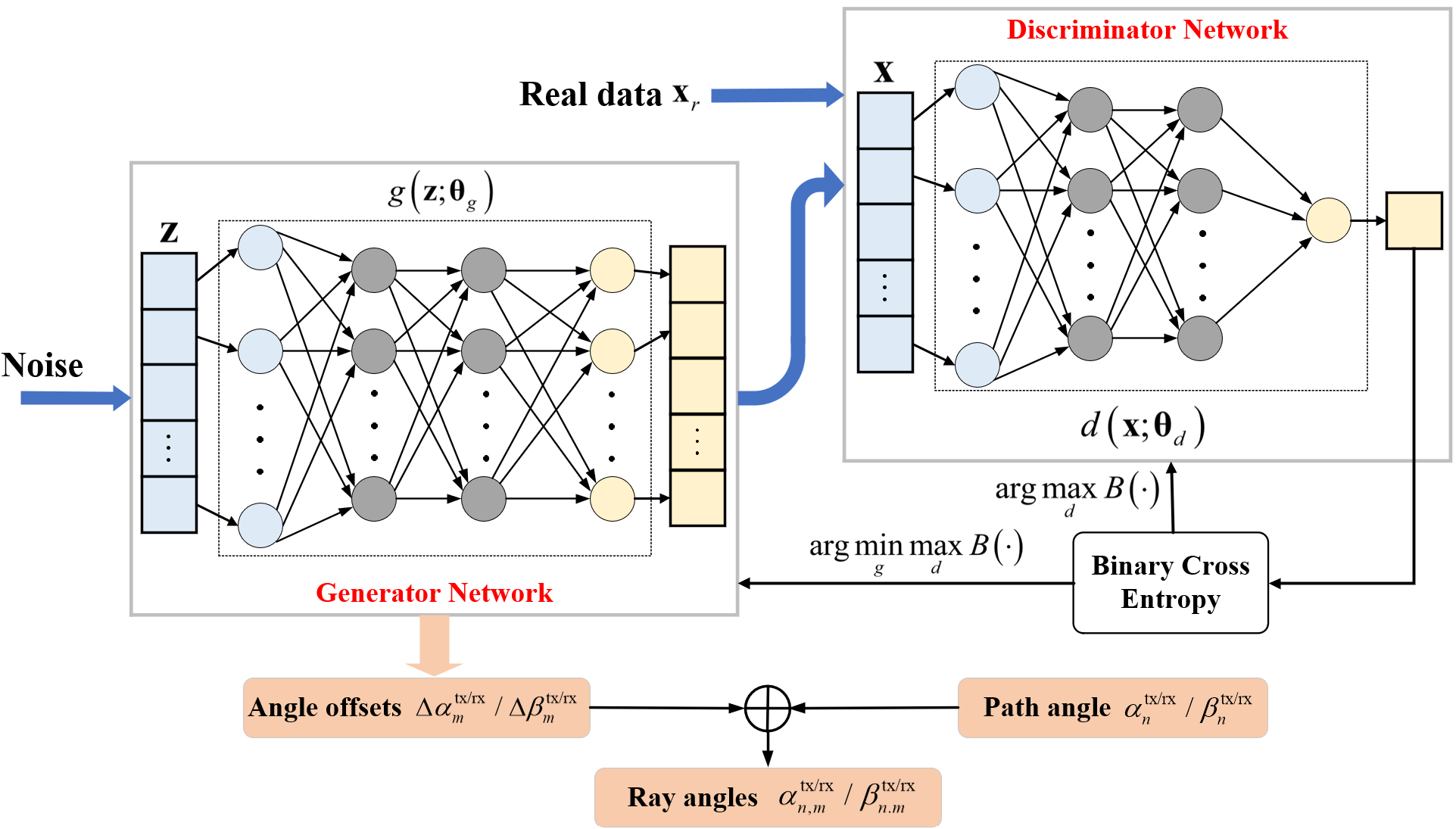}
	\caption{GAN-based generation framework of ray angles.}
    \label{fig:4}
\end{figure*}

\par Then, each ray angle can be generated in a stochastic way by adding random offsets on the path angle. We have conducted quantities of RT simulations for U2V mmWave channel at 28 GHz under urban, suburban, hilly, and sea scenarios. There are 25000 channels under each studied scenario and each channel contains a large number of propagation rays. In order to obtain the statistical properties of angle offsets, it is necessary to cluster the rays into different paths and remove the mean angle. The K-means method is widely used for clustering purposes, due to its effectiveness and robustness. However, the Euclidean distance normally applied in the traditional K-means algorithm is not suitable for channel clustering. Recently, the authors in \cite{HeR18_ICM, LiuL20_Access} used ray delay and angle as the Euclidean distance. However, the clustering performance is found to be significantly affected by the delay and angle weight settings in the simulation. Therefore, a modified Euclidean distance is applied in this paper as
\begin{equation}
D\left( {{c_i},{c_j}} \right) = \sqrt {{D_{{c_i}{\rm{, }}{c_j}}}\left( \tau  \right) + {D_{{c_i}{\rm{, }}{c_j}}}\left( \alpha  \right) + {D_{{c_i}{\rm{, }}{c_j}}}\left( \beta  \right)}
\label{23}
\end{equation}

\noindent where ${D_{{c_i}{\rm{, }}{c_j}}}\left( \tau  \right)$, ${D_{{c_i}{\rm{, }}{c_j}}}\left( \alpha  \right)$ and ${D_{{c_i}{\rm{, }}{c_j}}}\left( \beta  \right)$ are the distances with respect to delay space, azimuth angle space, and elevation angle space between two different rays respectively. They can be further expressed as
\begin{equation}
{D_{{c_i}{\rm{, }}{c_j}}}\left( x \right) = {\left( {{\xi _x}\left( {{x_i} - {x_j}} \right)} \right)^2}
\label{24}
\end{equation}

\noindent where $x \in \left\{ {\tau ,\alpha ,\beta } \right\}$ and ${\xi _x}$ is the weight coefficient of each parameter.
\par According to the modified distance, the procedure of proposed K-means clustering algorithm is shown in Algorithm~1, where ${{\bf{C}}_m} = \left\{ {{c_m} = \left( {{\tau _m},{\alpha _m},{\beta _m}} \right)\left| {m = 1,2,...,{M_k}} \right.} \right\}$ denotes the input ray with ${\tau _m}$, ${\alpha _m}$, ${\beta _m}$ representing the delay, AAoA, and EAoA of $m$th ray, ${N_k}$ is the cluster number and ${{\bf{C}}_n}$ is the output cluster aggregate. It should be mentioned that the input cluster number is difficult to manually choose from the massive random channel data. In this paper, an elbow method is developed to predict the optimal ${N_k}$. Firstly, the sum of squared errors (SSE) with respect to different ${N_k}$ is defined as
\begin{equation}
S\left( {{N_k}} \right) = \sum\limits_{i = 1}^{{N_k}} {\sum\limits_{{c_n} \in {{\bf{C}}_n}} {D\left( {{c_n},{u_i}} \right)} }
\label{25}
\end{equation}

\noindent where ${u_i}$ is the centroid of each cluster. Then, we can obtain an elbow curve by sampling the lower values of SSE. The first point on the elbow curve that lower than the SSE threshold and decreasing slope threshold could be found. In this paper, the elbow point is chosen as the optimal cluster number ${N_k}$.

\begin{table}[!bt]
\centering
\centering
\renewcommand\arraystretch{1.2}
\begin{tabular}{p{8cm}}
\hline
$\textbf{Algorithm 1}$:  K-means based clustering method \\ \hline
$\textbf{Input}$: Aggregate ${{\bf{C}}_m}$, ${N_k}$ ; \\
$\textbf{Output}$: Aggregate ${{\bf{C}}_n}$ ;\\
\ 1:$\textbf{Initial}$   $k = 1$ ; \\
\ 2:Choose ${N_k}$ initial centroids $u_1^{(1)},u_2^{(1)}, \cdots ,u_{{N_k}}^{(1)}$ ,\\
    define $u_i^{(1)} \ne u_i^{(0)},i \in [1\ {\rm{  }}{N_k}]$ ;\\
\ 3:$\textbf{while}$ $u_i^{(k)} \ne u_i^{(k - 1)}$ $\textbf{do}$ ; \\
\ 4:\ \ \ ${{\bf{C}}_i} = \emptyset ,i \in [1\ {\rm{  }}{N_k}]$ ;\\
\ 5:\ \ \ $\textbf{for}$ m = 1 :${M_k}$ $\textbf{do}$ \\
\ 6:\ \ \ \ ${d_{i,m}} = D\left( {u_i^{(k)},{c_m}} \right),i \in [1\ {\rm{ }}{N_k}]$; \\
\ 7:\ \ \ \ ${\lambda _m} = \mathop {\arg \min }\limits_{i \in [1\ {N_k}]} \left( {{d_{i,m}}} \right),{\lambda _m} \in [1\ {\rm{  }}{N_k}]$; \\
\ 8:\ \ \ \ ${{\bf{C}}_{{\lambda _m}}} = {{\bf{C}}_{{\lambda _m}}} \cup \left\{ {{c_m}} \right\}$; \\
\ 9:\ \ \ $\textbf{end for}$ \\
\ 10:\ \ \ \ k = k + 1; \\
\ 11:\ \ \ \ $\textbf{for}$  i = 1 :${N_k}$  $\textbf{do}$ \\
\ 12:\ \ \ \ \ $u_i^{(k)} = \frac{1}{{\left| {{{\bf{C}}_i}} \right|}}\sum\limits_{{c_i} \in {{\bf{C}}_i}} {{c_i}} $ ;\\
\ 13:\ \ \ $\textbf{end for}$\\
\ 14:$\textbf{end while}$\\
\ 15:${{\bf{C}}_n} = \left\{ {{{\bf{C}}_{{\lambda _m}}}|{\lambda _m} \in [1\ {\rm{  }}{N_k}]} \right\}$; \\
\hline
\end{tabular}
\label{table II}
\vspace{0cm}  
\setlength{\abovecaptionskip}{-0.2cm}   
\setlength{\belowcaptionskip}{0cm}   
\end{table}

\par Based on the clustering results, we can acquire massive angle offsets by removing the mean angle from each clustered rays. Most of previous methods including some standardized models, e.g., 3GPP and Winner, usually adopt the specific probability density function (PDF), i.e., Gaussian \cite{3GPP18, WINNER+}, Laplace\cite{WINNER+}, Von mise \cite{ZhuQ19_IET, ChengX20_ITJ, HeR20_Trans}, and Von mise Fisher \cite{ZhuQ19_WCL} to fit the angle offset distribution. Then, the angle offset following these PDFs can be generated randomly. This method is easy to apply but it fails when there is no suitable distribution to fit the measured results.
\par As an alternative generative network, GAN is recently applied to generate samples with the same distribution as the training data \cite{YangY19_ICM, O'Shea19_ICNC, HuTY21_Trans}. Since it does not need the information of underlying distribution, the shortcoming of traditional methods can be overcome. Our GAN-based generation framework of angle offsets is shown in Fig. 4. It mainly includes three parts, i.e., generator network (GN), discriminator network (DN), and cost function. In the GN, the neural network can be equivalent to any arbitrary distribution function $g\left( {{\bf{z}};{{\bm{\theta }}_g}} \right)$ with respect to ${{\bm{\theta }}_g}$. The neural network in DN can be equivalent to a quadratic classifier $d\left( {{\bf{x}};{{\bm{\theta }}_d}} \right)$ with respect to ${{\bm{\theta }}_d}$.
\par To train the equivalent parameters ${{\bm{\theta }}_g}$ and ${{\bm{\theta }}_d}$ efficiently, a cost function of binary cross entropy (BCE) is designed and it can be expressed as
\begin{equation}
\begin{aligned}
B\left( {{{\bm{\theta }}_g},{{\bm{\theta }}_d}} \right) &= {{\rm{E}}_{{\bf{x}} \subseteq {{\bf{x}}_r}}}\left[ {\log d\left( {{\bf{x}};{{\bm{\theta }}_d}} \right)} \right] \\
&\ \ \ + {{\rm{E}}_{{\bf{x}} \subseteq g\left( {{\bf{z}};{{\bm{\theta }}_g}} \right)}}\left[ {\log \left( {1 - d\left( {{\bf{x}};{{\bm{\theta }}_d}} \right)} \right)} \right]
\end{aligned}
\label{26}
\end{equation}
\noindent where ${\rm{E}}\left[  \cdot  \right]$ is the expectation function. During the training procedure, both the GN and DN are pursuing the highest benefit and the benefit functions of GN and DN can be defined respectively as
\begin{equation}
{d^*}\left( {{{\bm{\theta }}_d}} \right) = \arg \mathop {\max }\limits_d B\left( {{{\bm{\theta }}_g},{{\bm{\theta }}_d}} \right)
\label{27}
\end{equation}
\begin{equation}
{g^*}\left( {{{\bm{\theta }}_g}} \right) = \arg \mathop {\min }\limits_g \mathop {\max }\limits_d B\left( {{{\bm{\theta }}_g},{{\bm{\theta }}_d}} \right).
\label{28}
\end{equation}
\noindent Finally, the trained GNs can be used to generate the angle offsets.

\section{Theoretical results of statistical properties}
Due to the randomness of generated channels, it is impossible to verify the correctness of channel model by comparing the generated data with measured ones directly. Instead, the statistical properties of channels can be used to validate the effectiveness and correctness of the proposed modeling approach. The channel statistical properties are also important for the channel coding, channel estimation, interleave scheme, and so on in the communication system. In this following, the statistical properties, i.e., PDP, ACF, CCF, and DPSD of proposed U2V channel model are derived for theoretical comparison and validation purposes.
\par For the proposed U2V mmWave channel model, the channel transfer function can be obtained by using Fourier transform of (3) as
\begin{equation}
\begin{aligned}
{H_{qp}}\left( {f,t,{\bf{d}}} \right) &= \int_{ - \infty }^\infty  {{h_{qp}}\left( {t,\tau ;{\bf{w}},{\bf{b}},{\bm{\sigma }};{{\bm{\theta }}_g}} \right){{\rm{e}}^{ - {\rm{j}}2\pi f\tau }}} {\rm{d}}\tau \\
&= \sum\limits_{n = 1}^{N\left( t \right)} {\sqrt {{P_n}\left( {t;{\tau _n},{\bf{w}},{\bf{b}},{\bm{\sigma }}} \right)} {{\widetilde h}_n}\left( {t;{{\bm{\theta }}_g}} \right)} {{\rm{e}}^{ - {\rm{j}}2\pi f{\tau _n}\left( t \right)}}
\end{aligned}
\label{29}
\end{equation}

\noindent where ${\bf{d}} = \left\{ {{{\bf{d}}^{'}}^{\rm{tx}},{{\bf{d}}^{'}}^{\rm{rx}}} \right\}$ is the location vector of Tx and Rx antenna with ${{\bf{d}}^{'}}^{\rm{tx}}(t) = {\bf{R}}_{\bf{v}}^{{\rm{tx}}}(t) \cdot {{\bf{R}}^{\rm{P}}}(t) \cdot {{\bf{d}}^{{\rm{tx}}}}(t)$ and ${{\bf{d}}^{'}}^{\rm{rx}}(t) = {\bf{R}}_{\bf{v}}^{{\rm{rx}}}(t) \cdot {{\bf{d}}^{{\rm{rx}}}}(t)$. Note that ${{\bf{d}}^{'}}^{{\rm{tx}}}(t)$ is time-variant due to the movement and attitude variations of UAV and ${{\bf{d}}^{'}}^{{\rm{rx}}}(t)$ is time-variant due to the movement of vehicle. By setting $\Delta f = 0$ , the spatial-temporal correlation function (STCF) of proposed model can be defined as (31),
\begin{figure*}[!b]
\normalsize
\begin{equation}
\begin{array}{c}
{\rho _{{q_1}{p_1},}}_{{q_2}{p_2}}\left( {t,{\bf{d}};\Delta t,\Delta {\bf{d}}} \right) = {\rm{E}}\left\{ {\frac{{H_{{q_1}{p_1}}^*\left( {t,{\bf{d}}} \right){H_{{q_2}{p_2}}}\left( {t{\rm{ + }}\Delta t,{\bf{d}} + \Delta {\bf{d}}} \right)}}{{\left| {H_{{q_1}{p_1}}^*\left( {t,{\bf{d}}} \right)} \right|\left| {{H_{{q_2}{p_2}}}\left( {t{\rm{ + }}\Delta t,{\bf{d}} + \Delta {\bf{d}}} \right)} \right|}}} \right\}
\end{array}
\label{30}
\end{equation}
\end{figure*}
\begin{figure*}[!b]
\normalsize
\begin{equation}
\begin{aligned}
{P_{{h_{qp}}}}\left( {t,\tau ;{\bf{w}},{\bf{b}},{\bm{\sigma }};{{\bm{\theta }}_g}} \right) &= {\left| {{h_{qp}}\left( {t,\tau ;{\bf{w}},{\bf{b}},{\bm{\sigma }};{{\bm{\theta }}_g}} \right)} \right|^2}\\
 &= {h_{qp}}\left( {t,\tau ;{\bf{w}},{\bf{b}},{\bm{\sigma }};{{\bm{\theta }}_g}} \right) \cdot h_{qp}^*\left( {t,\tau ;{\bf{w}},{\bf{b}},{\bm{\sigma }};{{\bm{\theta }}_g}} \right)\\
 &= \sum\limits_{n = 1}^{N\left( t \right)} {\frac{{{P_n}\left( {t;{\tau _n},{\bf{w}},{\bf{b}},{\bm{\sigma }}} \right)}}{M}\sum\limits_{m = 1}^M {{e^{{\rm{j}}\psi \left( {t,{\bf{d}};{{\bm{\theta }}_g}} \right)}}} \sum\limits_{m = 1}^M {{e^{{\rm{ - j}}\psi \left( {t,{\bf{d}};{{\bm{\theta }}_g}} \right)}}} \delta \left( {\tau  - {\tau _n}\left( t \right)} \right)}
\end{aligned}
\label{31}
\end{equation}
\end{figure*}
\begin{figure*}[!b]
\normalsize
\begin{equation}
\tilde \rho _{qp}^n\left( {t,{\bf{d}};\Delta t;{{\bm{\theta }}_g}} \right) = {\rm{E}}\left\{ \begin{array}{l}
\sum\limits_{m = 1}^M {{e^{{\rm{ - j}}\psi \left( {t;{\bf{d}};{{\bm{\theta }}_g}} \right)}}}  \cdot \sum\limits_{m = 1}^M {{e^{{\rm{j}}\psi \left( {t + \Delta t;{\bf{d}};{{\bm{\theta }}_g}} \right)}}}
 \cdot {{\rm{e}}^{{\rm{ - j}}2\pi f\left( {{\tau _n}\left( t \right) - {\tau _n}\left( {t + \Delta t} \right)} \right)}} \\
 /\left| {\sum\limits_{m = 1}^M {{e^{{\rm{ - j}}\psi \left( {t;{\bf{d}};{{\bm{\theta }}_g}} \right)}}} } \right|/\left| {\sum\limits_{m = 1}^M {{e^{{\rm{j}}\psi \left( {t + \Delta t;{\bf{d}};{{\bm{\theta }}_g}} \right)}}} } \right|
\end{array} \right\}.
\tag{34}
\end{equation}
\end{figure*}
\begin{figure*}[!b]
\normalsize
\begin{equation}
\tilde \rho _{{q_1}{p_1},{q_2}{p_2}}^n\left( {t,{\bf{d}};\Delta {\bf{d}};{{\bm{\theta }}_g}} \right) = {\rm{E}}\left\{ \begin{array}{l}
\sum\limits_{m = 1}^M {{e^{{\rm{ - j}}\psi \left( {t;{\bf{d}};{{\bm{\theta }}_g}} \right)}}}  \cdot \sum\limits_{m = 1}^M {{e^{{\rm{j}}\psi \left( {t;{\bf{d}} + \Delta {\bf{d}};{{\bm{\theta }}_g}} \right)}}} \\
/\left| {\sum\limits_{m = 1}^M {{e^{{\rm{ - j}}\psi \left( {t;{\bf{d}};{{\bm{\theta }}_g}} \right)}}} } \right|/\left| {\sum\limits_{m = 1}^M {{e^{{\rm{j}}\psi \left( {t;{\bf{d}} + \Delta {\bf{d}};{{\bm{\theta }}_g}} \right)}}} } \right|
\end{array} \right\}.
\tag{35}
\end{equation}
\end{figure*}
\begin{figure*}[!b]
\normalsize
\begin{equation}
\begin{array}{c}
{S_{qp}}\left( {f;t;{\bf{w}},{\bf{b}},{\bm{\sigma }};{{\bm{\theta }}_g}} \right) = \left| {\sum\limits_{n = 1}^{N(t)} {\int_{ - \infty }^\infty  {\rho _{qp}^n\left( {t,{\bf{d}};\Delta t;{{\bm{\theta }}_g}} \right){e^{ - {\rm{j2\pi }}f\Delta t}}w(t - \Delta t){\rm{d}}\Delta t} } } \right|\\
 = \left| {\sum\limits_{n = 1}^{N(t)} {{P_n}\left( {t;{\tau _n},{\bf{w}},{\bf{b}},{\bm{\sigma }}} \right)\int_{ - \infty }^\infty  {\left\{ {\begin{array}{*{20}{c}}
{\sum\limits_{m = 1}^M {{e^{{\rm{ - j}}\psi \left( {t;{\bf{d}};{{\bm{\theta }}_g}} \right)}}}  \cdot \sum\limits_{m = 1}^M {{e^{{\rm{j}}\psi \left( {t + \Delta t;{\bf{d}};{{\bm{\theta }}_g}} \right)}}} }\\
{ \cdot {{\rm{e}}^{{\rm{ - j}}2\pi f\left( {{\tau _n}\left( t \right) - {\tau _n}\left( {t + \Delta t} \right) - \Delta t} \right)}}w(t - \Delta t)}
\end{array}} \right\}{\rm{d}}\Delta t} } } \right|
\end{array}
\tag{36}
\end{equation}
\end{figure*}
\noindent where ${\left(  \cdot  \right)^ * }$ denotes the complex conjugation operation, $\Delta {\bf{d}} = \left\{ {\Delta {{\bf{d}}^{\rm{tx}}},\Delta {{\bf{d}}^{\rm{rx}}}} \right\}$ is the space lag (or antenna distance) at the Tx and Rx with $\Delta {{\bf{d}}^{\rm{tx}}} = {{\bf{d}}^{'}}_{{q_2}}^{\rm{tx}} - {{\bf{d}}^{'}}_{{q_1}}^{\rm{tx}}$ and $\Delta {{\bf{d}}^{\rm{rx}}} = {{\bf{d}}^{'}}_{{p_2}}^{\rm{rx}} - {{\bf{d}}^{'}}_{{p_1}}^{\rm{rx}}$.

\subsection{Time-variant PDP}
The PDP is an important characteristic to describe the channel in delay domain. It demonstrates the path power distribution of received signal with respect to different delays. The PDP of proposed model can be defined as (32),
\noindent where the time-variant phase $\psi \left( {t;{{\bm{\theta }}_g}} \right)$ is rewritten by $\psi \left( {t,{\bf{d}};{{\bm{\theta }}_g}} \right)$ for the convenience of describing STCF. Considering that the mean power of propagation rays within $n$th path is approximately equals to path power $P_n$, we can obtain $\left| {\sum\limits_{m = 1}^M {{e^{{\rm{j}}\psi \left( {t,{\bf{d}};{{\bm{\theta }}_g}} \right)}}} } \right| \approx \sqrt M$. Since the PDP only concerns the power and the channel phases do not affect the power a lot, we can simplify (32) as
\begin{equation}
{P_{{h_{qp}}}}\left( {t,\tau ;{\bf{w}},{\bf{b}},{\bm{\sigma }}} \right) = \sum\limits_{n = 1}^{N\left( t \right)} {{P_n}\left( {t;{\tau _n},{\bf{w}},{\bf{b}},{\bm{\sigma }}} \right)\delta \left( {\tau  - {\tau _n}\left( t \right)} \right)} .
\tag{33}
\end{equation}
\vspace{-0.5cm}  

\subsection{Time-variant ACF and CCF}
Due to the insufficient time interval or antenna interval, the received signal have certain correlation. For example, ACF is used to express the correlation at different time instants for the same channel. By substituting ${q_1}{\rm{ = }}{q_2}{\rm{ = }}q$, and ${p_1}{\rm{ = }}{p_2}{\rm{ = }}p$ into (31), i.e., $\Delta {\bf{d}} = 0$, the normalized ACF of $n$th path can be expressed as (34).
\noindent On the other hand, CCF denotes the correlation for different channels at the same time instant. By substituting $\Delta t{\rm{ = }}0$ into (31), the normalized CCF of $n$th path can be obtained as (35).
\noindent Note that $\Delta {\bf{d}}$ is relative to the UAV rotation, so the attitude variation would have significant effect on the CCF.

\subsection{Time-variant DPSD}
The DPSD is an important characteristic of mobile channels, which reflects the power distribution of received signal in Doppler frequency domain. Since the U2V channel is non-stationary, the short-time Fourier transform should be used to calculate the DPSD of proposed model as (36),
\noindent where ${\rm{w}}\left( {t{\rm{ }} - {\rm{ }}\Delta t} \right)$ is the time window function in which the channel can be approximated stationary. It should be mentioned that the original ACF is used here instead of the normalized ACF.

\begin{figure*}[!b]
\centering
\subfigure[] {\includegraphics[width=85mm]{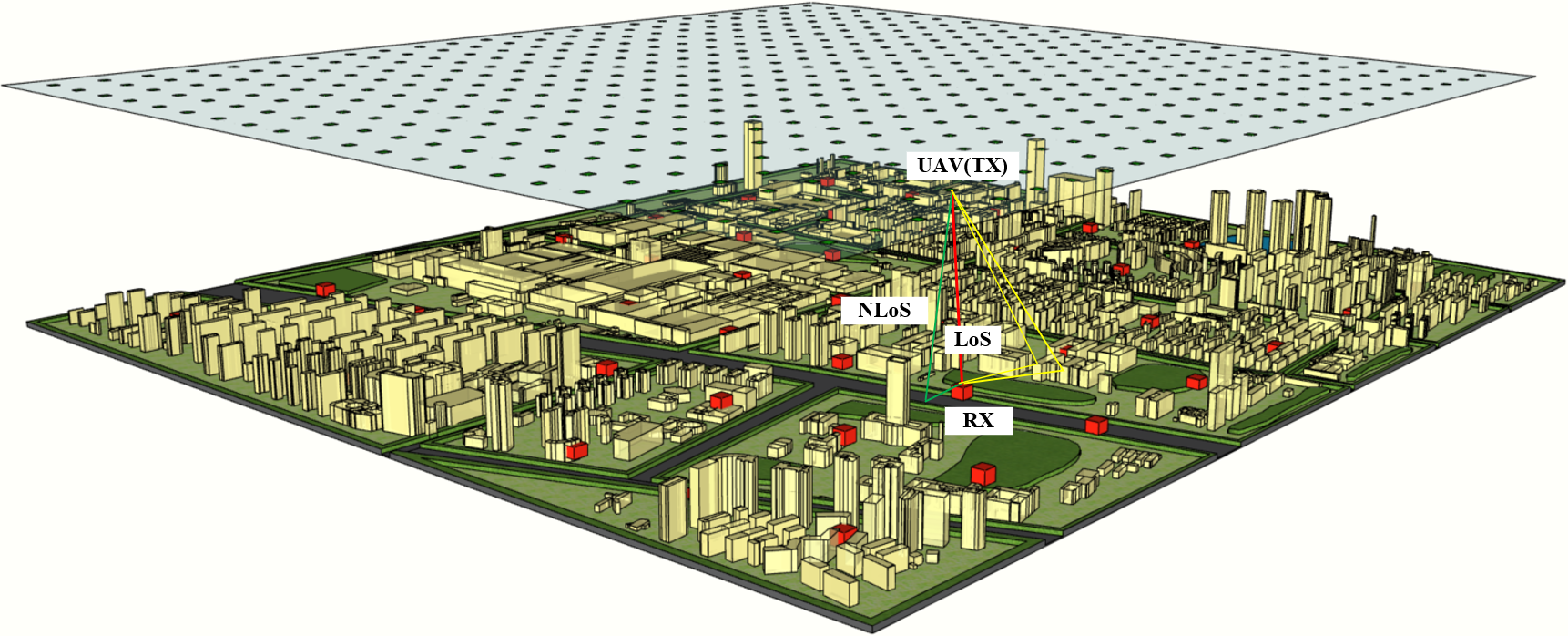}}
\subfigure[] {\includegraphics[width=85mm]{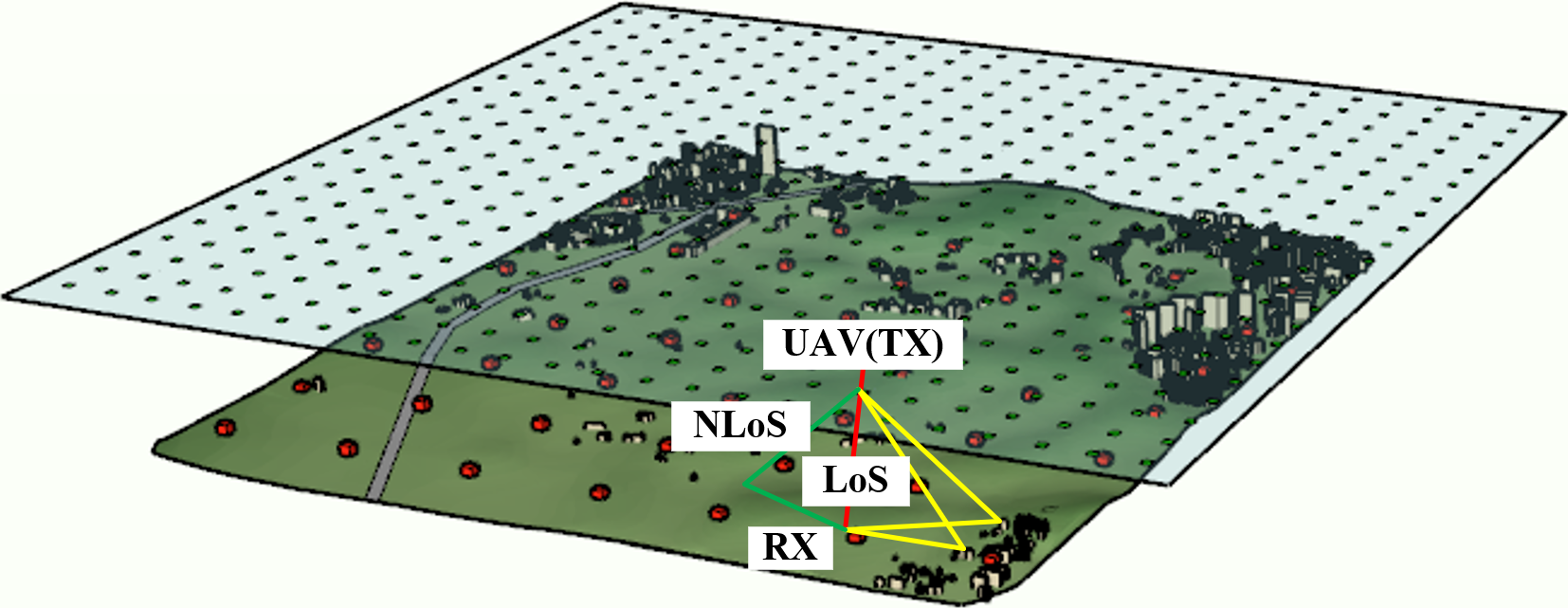}}
\subfigure[] {\includegraphics[width=85mm]{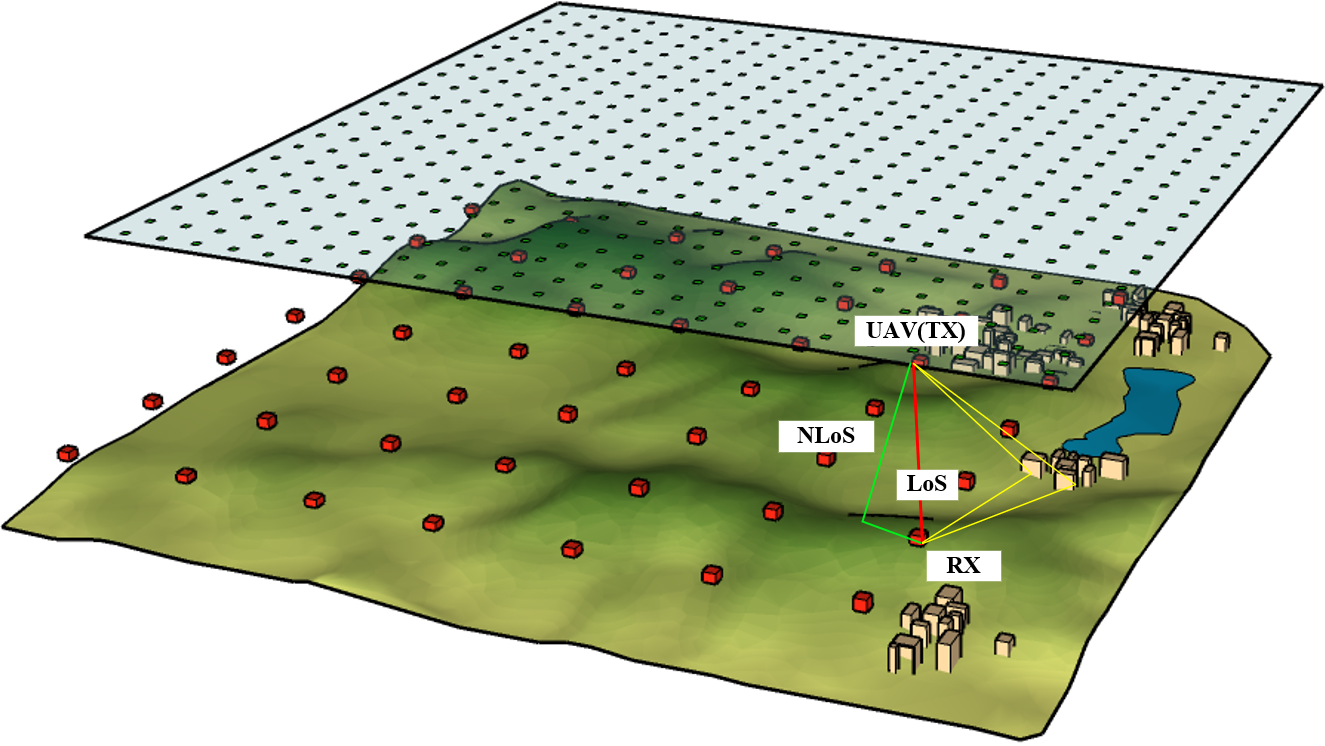}}
\subfigure[] {\includegraphics[width=85mm]{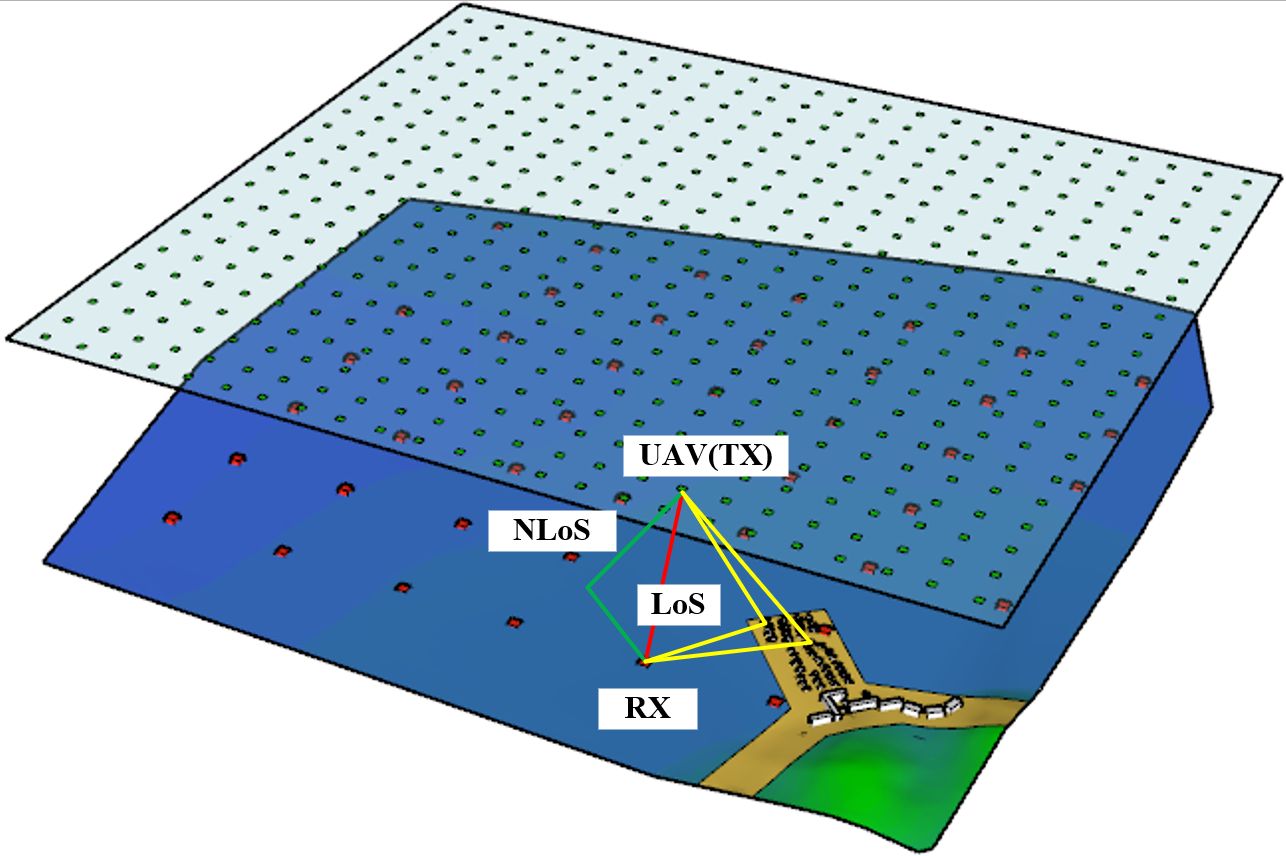}}
\caption{The acquisition of RT-based data set under (a) urban, (b) suburban, (c) hilly, and (d) sea scenarios.}
\label{fig:5}
\end{figure*}

\section{Simulation results and validation}
\subsection{Validation of parameter generation method}
\par To verify the effectiveness of proposed parameter generation method, we use the RT-based channel data to train and validate the ML-based parameter generation network. The acquisition of RT-based data set under four different scenarios, i.e., urban, suburban, hilly, and sea, are shown in Fig. 5. The urban, suburban, and hilly scenarios are located at Nanjing city, China and the sea scenario is located at Qingdao city, China. In each scenario, we set up 50 Rxs at the height of 2 m and 500 Txs at the height of 150 m. By applying the RT method, the U2V channel data between 25000 different pairs of transceivers are obtained in different scenarios at 28 GHz, where each channel contains massive sets of ray delay, power, and angle.
\par For the BPNN-based power generation network, taking the urban scenario as an example, the neuron number of hidden layer is set as 4 and the trained parameters of neurons are listed in Table II. It should be mentioned that in this simulation the relationship between path power and delay is not very complicated, so a small size BPNN is applied and the L2 regularization in the cost function is ignored as well. The sigmoid function is applied in hidden layers, since the layer number is small and the output layer does not use the activation function. In the neural network, the optimizer of ADAM is adopted and the learning rate is set as 0.001. Based on the trained network, the comparison of generated and original values is shown in Fig. 6. Note that LoS path and NLoS path are studied separately since the LoS path is not affected by the scattering propagation like NLoS paths. In Fig. 6, it can be found that the neural network is well trained and the prediction values are well consistent with the original ones in both the training set and validation set, which proves the prediction framework does not occur overfitting or underfitting problem and can describe the relationship between power and delay well. For comparison purpose, the exponential model for NLoS case in 3GPP is also referred to fit the training data as shown in Fig. 6. Note that the exponential function comes to linear with the power unit in dB and it cannot fit the generated values reliably as our proposed method.

\begin{table}[!tb]
\centering
\renewcommand\arraystretch{1.3}
\caption{The trained parameters of the neurons}
\centering
\begin{tabular}{|p{0.9cm}<{\centering}|p{0.9cm}<{\centering}|p{1cm}<{\centering}|p{1cm}<{\centering}|p{1.2cm}<{\centering}|p{1.2cm}<{\centering}|}
\hline
\multicolumn{2}{|c|}{Layer} & \multicolumn{2}{|c|}{hidden layer} & \multicolumn{2}{|c|}{output layer}  \\ \hline
\multirow{8}{1cm}{weight} & \multirow{4}{0.5cm}{LoS} & $w_{11}^{\left( 1 \right)}$ & -2.033 & $w_{11}^{\left( 2 \right)}$ & -0.343 \\ \cline{3-6}
 \multirow{8}{1cm}{} & \multirow{4}{1cm}{} & $w_{12}^{\left( 1 \right)}$ & 0.615 & $w_{21}^{\left( 2 \right)}$ & -21.993\\ \cline{3-6}
 \multirow{8}{1cm}{} & \multirow{4}{1cm}{}& $w_{13}^{\left( 1 \right)}$ & 1.951 & $w_{31}^{\left( 2 \right)}$ & -29.861\\ \cline{3-6}
 \multirow{8}{1cm}{} & \multirow{4}{1cm}{}& $w_{14}^{\left( 1 \right)}$ & 1.044 & $w_{41}^{\left( 2 \right)}$ & -29.012\\ \cline{2-6}

\multirow{8}{0.5cm}{} & \multirow{4}{0.8cm}{NLoS} & $w_{11}^{\left( 1 \right)}$ & 0.430 & $w_{11}^{\left( 2 \right)}$ & -18.140 \\ \cline{3-6}
 \multirow{8}{0.5cm}{} & \multirow{4}{1cm}{}& $w_{12}^{\left( 1 \right)}$ & 0.746 & $w_{21}^{\left( 2 \right)}$ & -32.827\\ \cline{3-6}
 \multirow{8}{0.5cm}{} & \multirow{4}{1cm}{}& $w_{13}^{\left( 1 \right)}$ & 0.772 & $w_{31}^{\left( 2 \right)}$ & -32.964\\ \cline{3-6}
  \multirow{8}{0.5cm}{} & \multirow{4}{1cm}{}& $w_{14}^{\left( 1 \right)}$ & 1.213 & $w_{41}^{\left( 2 \right)}$ & -32.980\\
\cline{1-6}

\multirow{8}{0.8cm}{bias} & \multirow{4}{0.5cm}{LoS} & $b_1^{\left( 1 \right)}$ & 0.050 & \multirow{4}{0.6cm}{$b_1^{\left( 2 \right)}$} & \multirow{4}{1.2cm}{-30.082} \\ \cline{3-4}
 \multirow{8}{1cm}{} & \multirow{4}{1cm}{} & $b_2^{\left( 1 \right)}$ & -1.323 & \multirow{4}{1cm}{} & \multirow{4}{1cm}{}\\ \cline{3-4}
 \multirow{8}{1cm}{} & \multirow{4}{1cm}{}& $b_3^{\left( 1 \right)}$ & 0.818 & \multirow{4}{1cm}{} & \multirow{4}{1cm}{}\\ \cline{3-4}
\multirow{8}{1cm}{} & \multirow{4}{1cm}{} & $b_4^{\left( 1 \right)}$ & 0.950 & \multirow{4}{1cm}{} & \multirow{4}{1cm}{}\\ \cline{2-6}

\multirow{8}{0.8cm}{} & \multirow{4}{0.8cm}{NLoS} & $b_1^{\left( 1 \right)}$ & -1.729 & \multirow{4}{0.6cm}{$b_1^{\left( 2 \right)}$} & \multirow{4}{1.2cm}{-33.212} \\ \cline{3-4}
\multirow{8}{1cm}{} & \multirow{4}{1cm}{} & $b_2^{\left( 1 \right)}$ & 0.922 & \multirow{4}{1cm}{} & \multirow{4}{1cm}{}\\ \cline{3-4}
\multirow{8}{1cm}{} & \multirow{4}{1cm}{} & $b_3^{\left( 1 \right)}$ & 1.033 & \multirow{4}{1cm}{} & \multirow{4}{1cm}{}\\ \cline{3-4}
\multirow{8}{1cm}{} & \multirow{4}{1cm}{} & $b_4^{\left( 1 \right)}$ & 1.345 & \multirow{4}{1cm}{} & \multirow{4}{1cm}{}\\ \cline{3-4}
\hline
\end{tabular}
\label{table II}
\end{table}

\vspace{-0.1cm}  
\begin{figure}[!b]
	\centering
	\includegraphics[width=88mm]{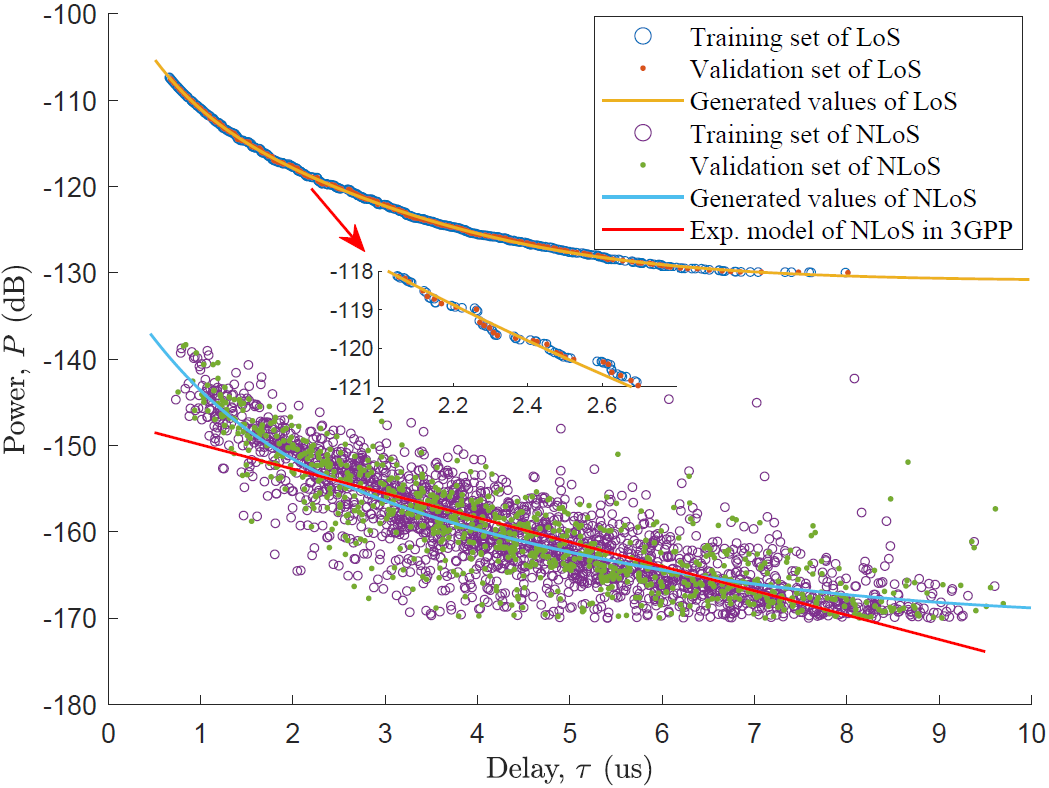}
	\caption{Training results for LoS path and NLoS path.}
    \label{fig:6}
\end{figure}

\par Similarly, based on the U2V channel data of urban scenario, the clustering results of AAoAs and EAoAs are shown in Fig. 7, respectively. As we can see in Fig. 7(a) that the elbow curve decreases with increasing cluster number. In this paper, we set the SSE threshold as 0.15 to ensure appropriate clustering precision and set the decreasing slope threshold as 0.005 to avoid over-clustering. Fig. 7(a) shows that the value of SSE is 0.11 and the decreasing slope of elbow curve is 0.0038 when  ${N_k} = 17$. Both of them are lower than their thresholds, respectively. Therefore, ${N_k} = 17$ is the optimal cluster number for this case. Meanwhile, Fig. 7(b) gives the clustered results of AAoAs and EAoAs, where marks with the same color belong to the same cluster. In the figure, the rays for this channel are well clustered into 17 valid paths and the value of normalized SSE is only 0.11, which validates the effectiveness of proposed clustering algorithm. Since only LoS and single-bounce paths are considered in this paper, the clustered results of AoD can be theoretically calculated from the knowledge of AoA and the environment geometry. Then we can acquire massive angle offsets by removing the mean angle from each clustered rays.

\begin{figure}[!tb]
\centering
\subfigure[] {\includegraphics[width=80mm]{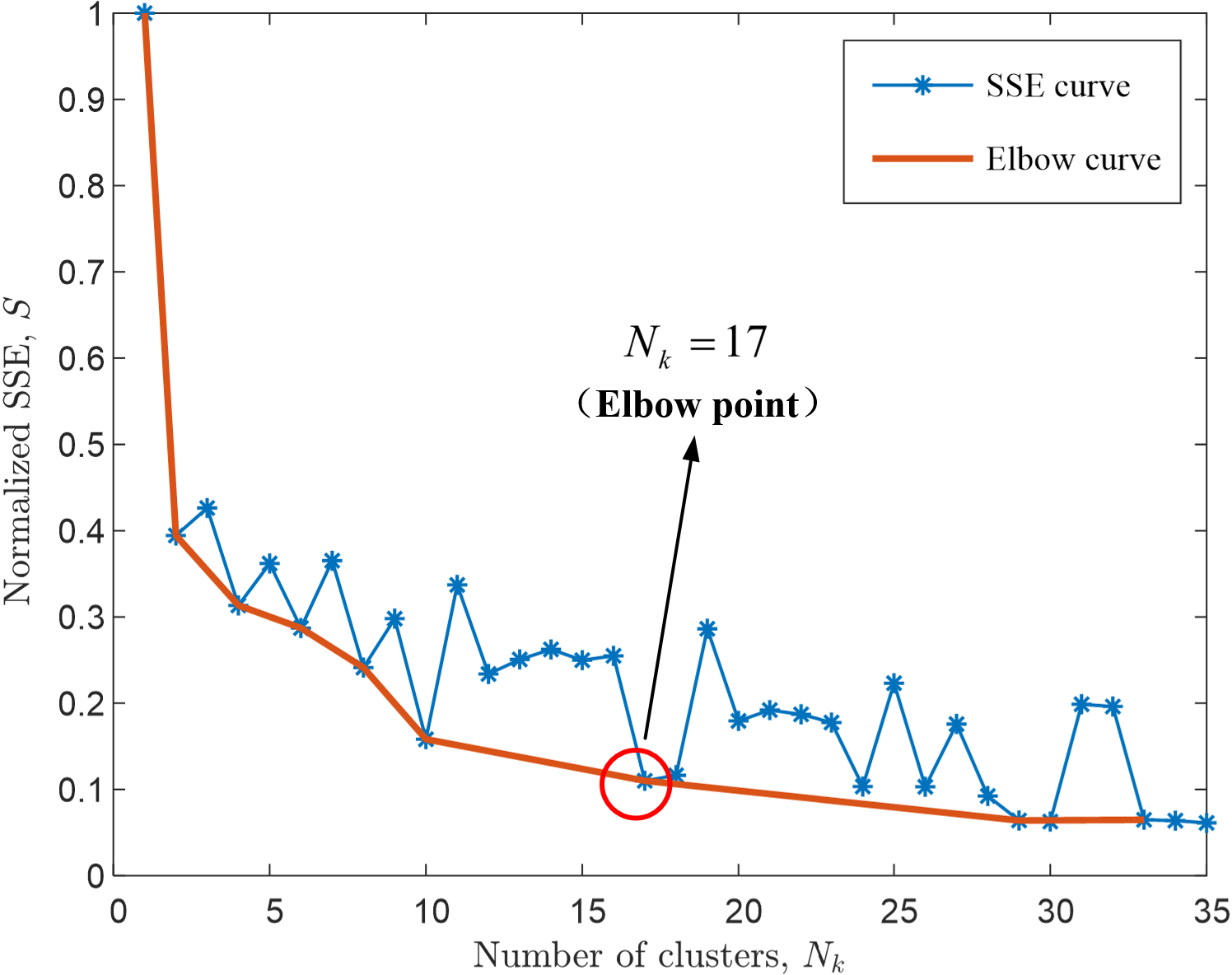}}
\subfigure[] {\includegraphics[width=85mm]{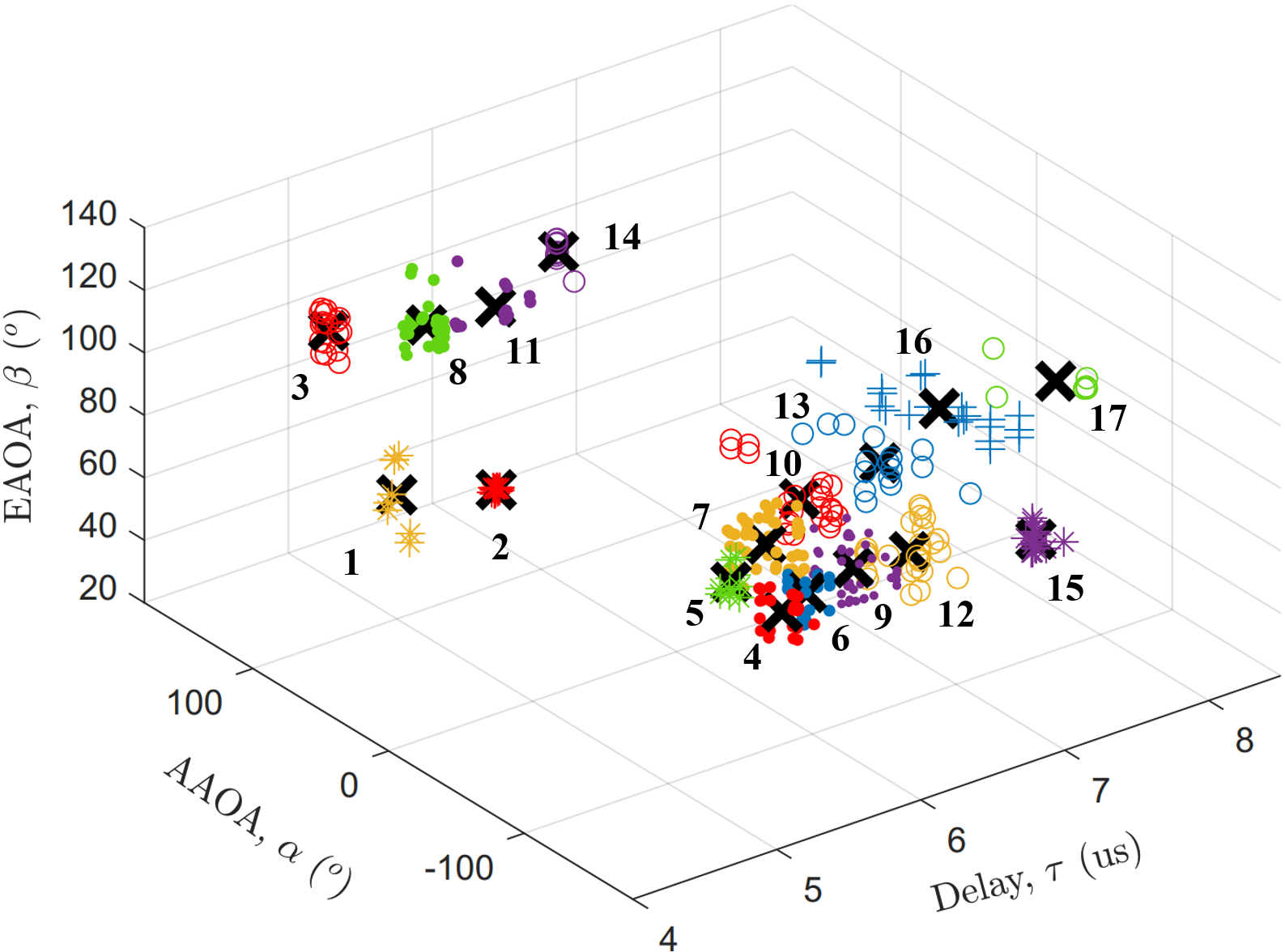}}
\caption{ (a) Optimal cluster number and (b) Clustering results of U2V mmWave channel under urban scenario.  }
\label{fig:7}
\end{figure}

\par According to the real data of angle offsets, the GANs for azimuth angle and elevation angle are trained, respectively. Take the U2V channel data under urban scenario as an example, the generated PDF is compared with the real one as well as the Gaussian distribution (GD) and Laplacian distribution (LD) studied in \cite{3GPP18} as shown in Fig. 8. Note that the conventional GD and LD for angles are obtained by using empirical expressions to approximately fit the channel data. However, it is difficult to manually find an accurate expression from massive channel data with limited knowledge and expertise of radio propagation. Therefore, there would be some deviations between the empirical expressions and real inner relationship of channel data. It can be seen from Fig. 8 that the generated one by GAN is well consistent with the real one and outperforms the conventional methods.

\vspace{-0.2cm}  
\begin{figure}[!t]
	\centering
	\includegraphics[width=85mm]{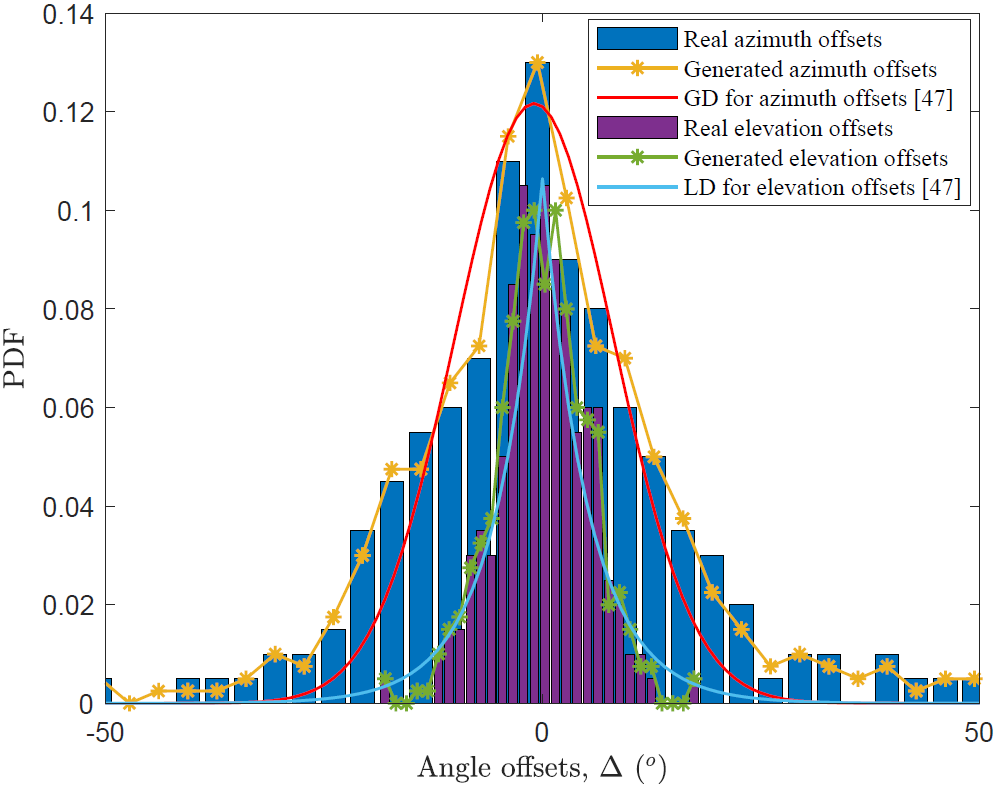}
	\caption{PDFs of training data, generated angle offsets and conventional methods.}
    \label{fig:8}
\end{figure}

\subsection{Validation of proposed channel model}
In this section, we use the RT and measured data to validate the proposed channel model. The U2V mmWave channel is studied under a typical urban scenario at 28 GHz. The scenario and trajectories with velocity and attitude variations are shown in Fig. 9. The UAV flies along a 3D trajectory with the velocity of 30 m/s at the height from 50 m to 200 m, and the vehicle moves along a straight road with the velocity of 5 m/s. It should be mentioned that the 3D attitude of UAV is also considered in the validation.

\begin{figure}[!b]
	\centering
	\includegraphics[width=75mm]{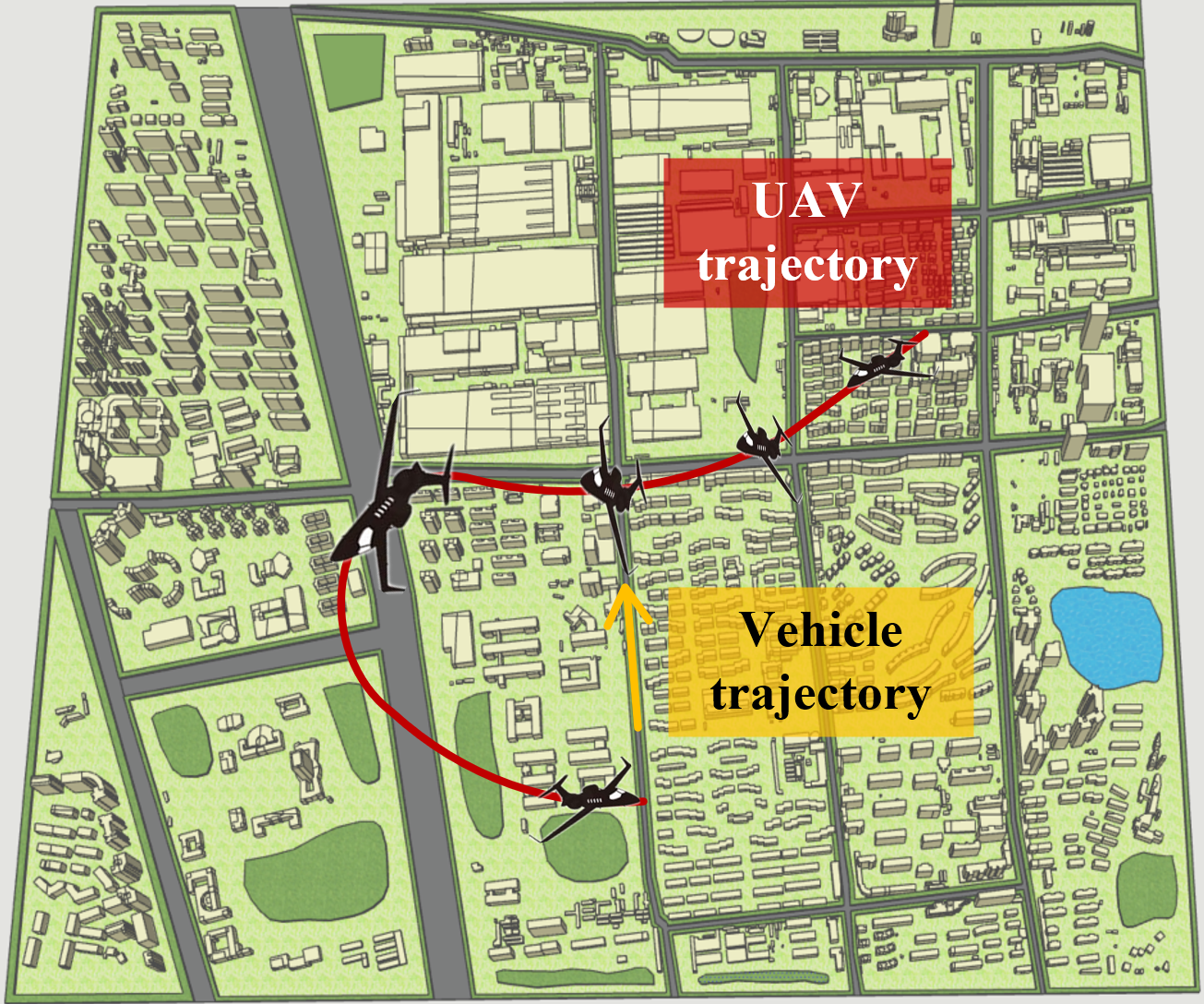}
	\caption{The trajectories of UAV with time-variant rotations and vehicle.}
    \label{fig:9}
\end{figure}

\par Based on the proposed generation method, the PDP of generated U2V channel is calculated by (33) and shown in Fig. 10(a). In the figure, the power and delay of LoS path in dark red is changing with the distance varying between the transceivers. Note that some NLoS paths in light red or green exists in the proximity of the LoS path which is likely to be the ground reflection path. For comparison purpose, the PDP under the same scenario is analyzed by RT method and given in Fig. 10(b). For the LoS path, the path delay of both proposed channel model and RT method varies between 0.5$\sim$2.7 us along with the time-variant distance of transceivers. Moreover, the LoS path almost disappears at 40$\sim$55 s due to the obstacle of scatterers. For the NLoS path, the path delay mainly distributes between 0.55$\sim$10 us. It can be found that the RT-based PDP provides more details for the hundreds of rays. The main reason is that it is a deterministic method and calculates all rays of each scatterer independently. However, for reducing the complexity, our proposed model only keeps the main information of scatterers such as the distribution and the adjacent rays are modeling as one cluster. Anyway, the PDPs of proposed method and RT method are showing good agreement in tendency and pattern. Moreover, both of them show the birth-death phenomenon of propagation paths, which has been observed in the field measurements \cite{SunR17_Trans}.
\par In order to validate the channel characteristic in the frequency domain, the time-variant DPSDs of proposed channel model and RT method are given in Fig. 11. In Fig. 11, we can see that the DPSD trend of proposed model is well consistent with the RT-based one under the urban scenario. At the beginning of LoS path, the UAV flies toward the vehicle and the relative angle between LoS path and velocity is about 30 degrees. We can calculate the theoretical value of Doppler frequency is about 2829 Hz. With the increasement of UAV height, the elevation angle increases and the Doppler frequency decreases. After $t$ = 28 s, the UAV flies away from the vehicle, and the Doppler frequency turns into negative. For the NLoS paths, the deterministic part (or the main trend) of proposed model is similar with the RT-based one. The Doppler frequencies distribute between -2800 Hz$\sim$2800 Hz, since each NLoS path has different relative angles between propagation path and velocity. For the previous stochastic UAV channel models \cite{ChengX19_IET, ZhangZC19_Access}, they were able to get the similar Doppler frequency of LoS path. However, in these models the scatterers were generated randomly and the birth-death phenomenon was ignored or the birth-death parameter was set as the recommended values, which is not well accorded with the specific propagation scenario.
\par Note that our proposed method consumes much less time than the RT method by generating the intra-path rays in a stochastic way. To verify the generation efficiency, the computational speed is compared under the scenario of Fig. 9 on the computer with Intel Xeon CPU E5 and 16 GB RAM. Firstly, both of them consume 39.6 s to deal with the digital map. Taking a 10-cluster channel with 120 rays as an example, the RT method consumes 9.4 s to track and calculate all the parameters of 120 rays. However, the new method only takes 1.4 s to obtain the inter-path parameters for 10 paths (or clusters) and takes tens of milliseconds to generate the ray parameters. If the channel state is updated per 100 ms, the RT method and our proposed method consume 979.6 s and 180.6 s for a 10-second simulation, respectively.

\begin{figure}[!tb]
\centering
\subfigure[] {\includegraphics[width=85mm]{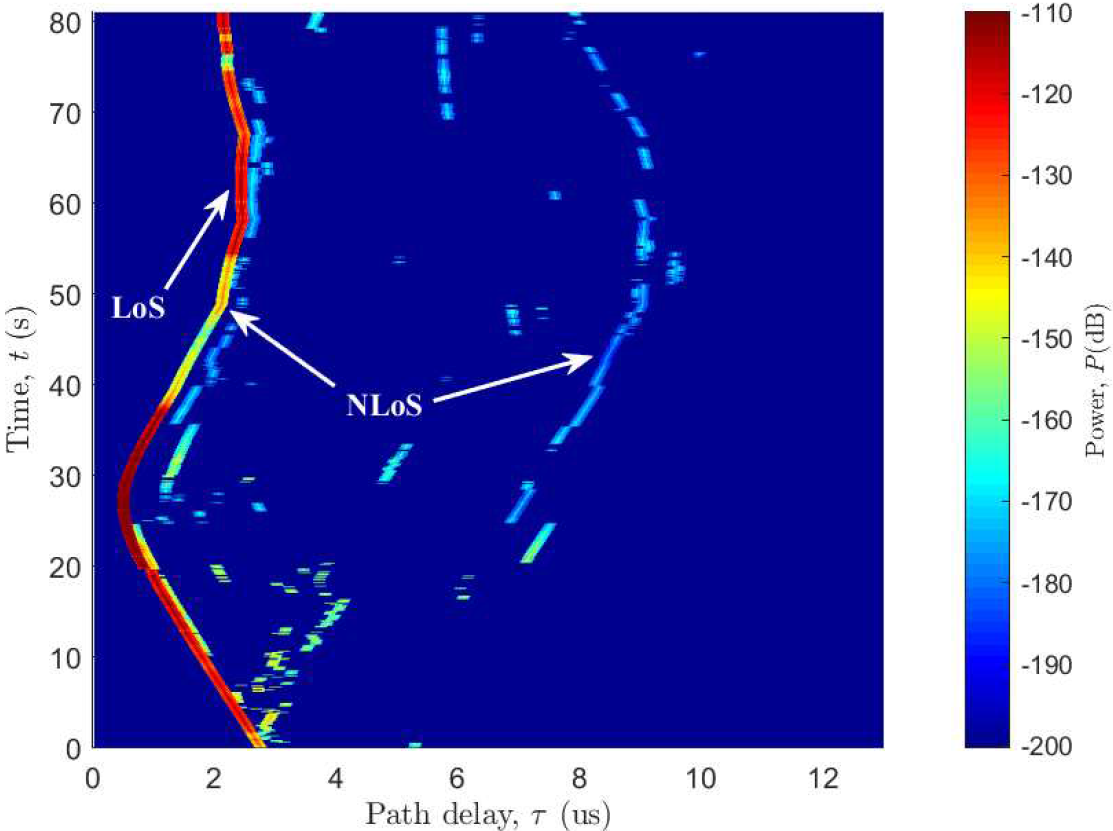}}
\subfigure[] {\includegraphics[width=85mm]{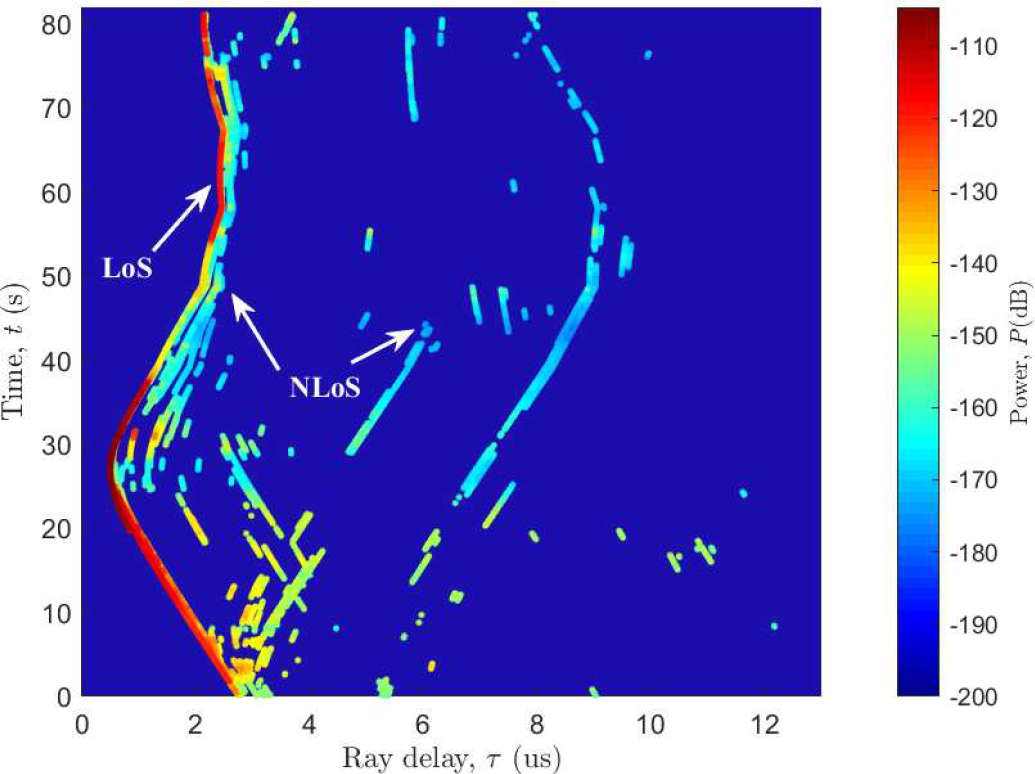}}
\caption{The time-variant PDPs of (a) proposed model and (b) RT method (urban scenario, ${f_0}$ = 28 GHz, $\left\| {{{\bf{v}}^{{\rm{tx}}}}} \right\|$ = 30 m/s, $\left\| {{{\bf{v}}^{{\rm{rx}}}}} \right\|$ = 5 m/s). }
\label{fig:9}
\end{figure}

\begin{figure}[!t]
\centering
\subfigure[] {\includegraphics[width=85mm]{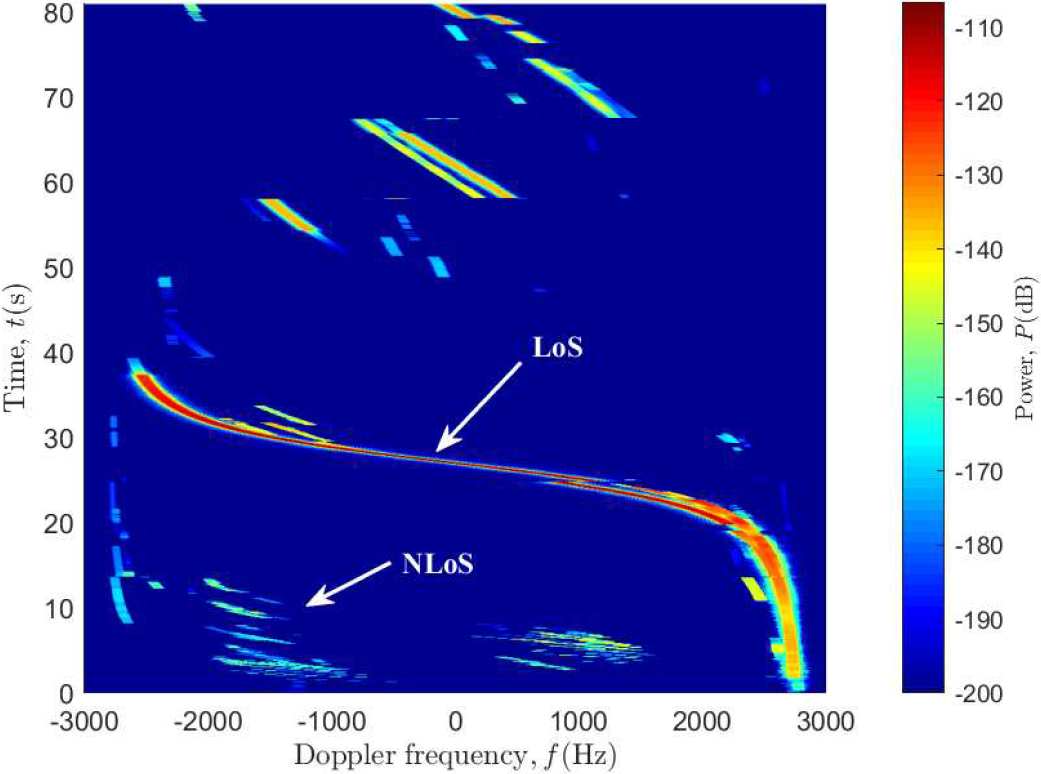}}
\subfigure[] {\includegraphics[width=85mm]{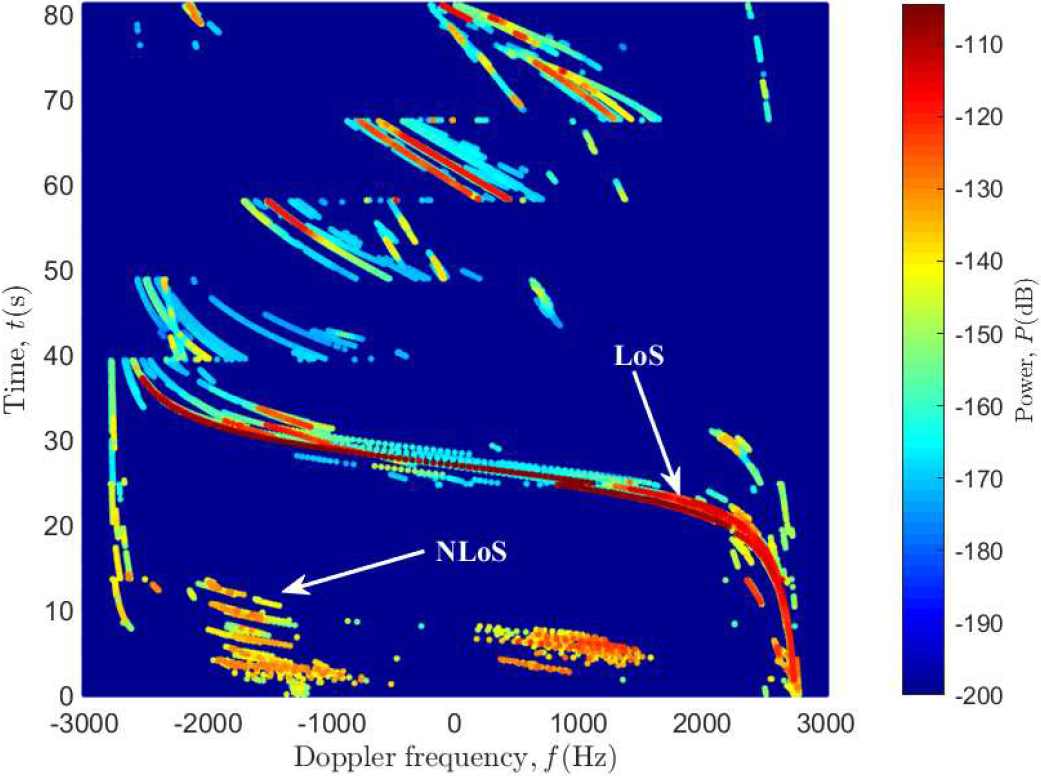}}
\caption{The time-variant DPSDs of (a) proposed model and (b) RT method (urban scenario, ${f_0}$ = 28 GHz, $\left\| {{{\bf{v}}^{{\rm{tx}}}}} \right\|$ = 30 m/s, $\left\| {{{\bf{v}}^{{\rm{rx}}}}} \right\|$ = 5 m/s). }
\label{fig:10}
\end{figure}

\par The attitude variation is a new feature of UAVs compared with traditional mobile terminals, which is included in the proposed model on the channel characteristic. In Section IV, we have derived the expression of CCF incorporating the factor of attitude variation. By substituting the simulation parameters into (35), the CCFs with/without attitude variation at three different time instants, i.e., $t$ = 0 s, $t$ = 25 s, and $t$ = 50 s are shown in Fig. 12. Noted that the antenna elements are assumed to be distributed on the y-axis, so the pitch angle would affect the antenna distance vector. For three time instants, the UAV is swerving with pitch attitude and the pitch angles are 0, 45, and 90 degrees, respectively. The rotation matrices caused by the attitude variation can be calculated as

\begin{equation}
\begin{array}{l}
{{\bf{R}}^{\rm{P}}}{|_{t = 0s}} = \left[ {\begin{array}{*{20}{c}}
1&0&0\\
0&1&0\\
0&0&1
\end{array}} \right]\\

{{\bf{R}}^{\rm{P}}}{|_{t = 25s}} = \left[ {\begin{array}{*{20}{c}}
1&0&0\\
0&1&0\\
0&{0.7071}&{0.7071}
\end{array}} \right]\\

{{\bf{R}}^{\rm{P}}}{|_{t = 50s}} = \left[ {\begin{array}{*{20}{c}}
1&0&0\\
0&1&0\\
0&1&0
\end{array}} \right].
\end{array}
\tag{37}
\end{equation}

In Fig. 12, the simulated results without attitude variation equals to the one with ${{\bf{R}}^{\rm{P}}}$ being identity matrix. Moreover, the results with attitude variation are obtained by substituting the generated channel data into the definition of CCF. It can be seen that the simulated results with and without attitude variation at $t$ = 0 s (LoS case) are the same since the pitch angle is $0^\circ$ at that time. The normalized spacing lag is around 0.13 at $70\%$ of the maximum CCF. When the UAV experiences the attitude variation at $t$ = 25 s (LoS case) and $t$ = 50 s (NLoS case), the simulated result with attitude variation decreases a little faster than the one without attitude variation in the pitch angle of 45 degrees, and this trend increases in the pitch angle of 90 degrees. At $t$ = 50 s, we can find that the normalized spacing lags with and without attitude variation are 0.09 and 0.11 at $50\%$ of the maximum CCF, respectively. For the proposed model, the simulated results are well consistent with the theoretical ones which validates the correctness of both theoretical derivation and generation method. Moreover, the CCF decreases to a low value when the normalized spacing lag is around 0.2, which could be a reference value for the antenna design of UAV mmWave communication system at 28 GHz.
\begin{figure}[!tb]
	\centering
	\includegraphics[width=80mm]{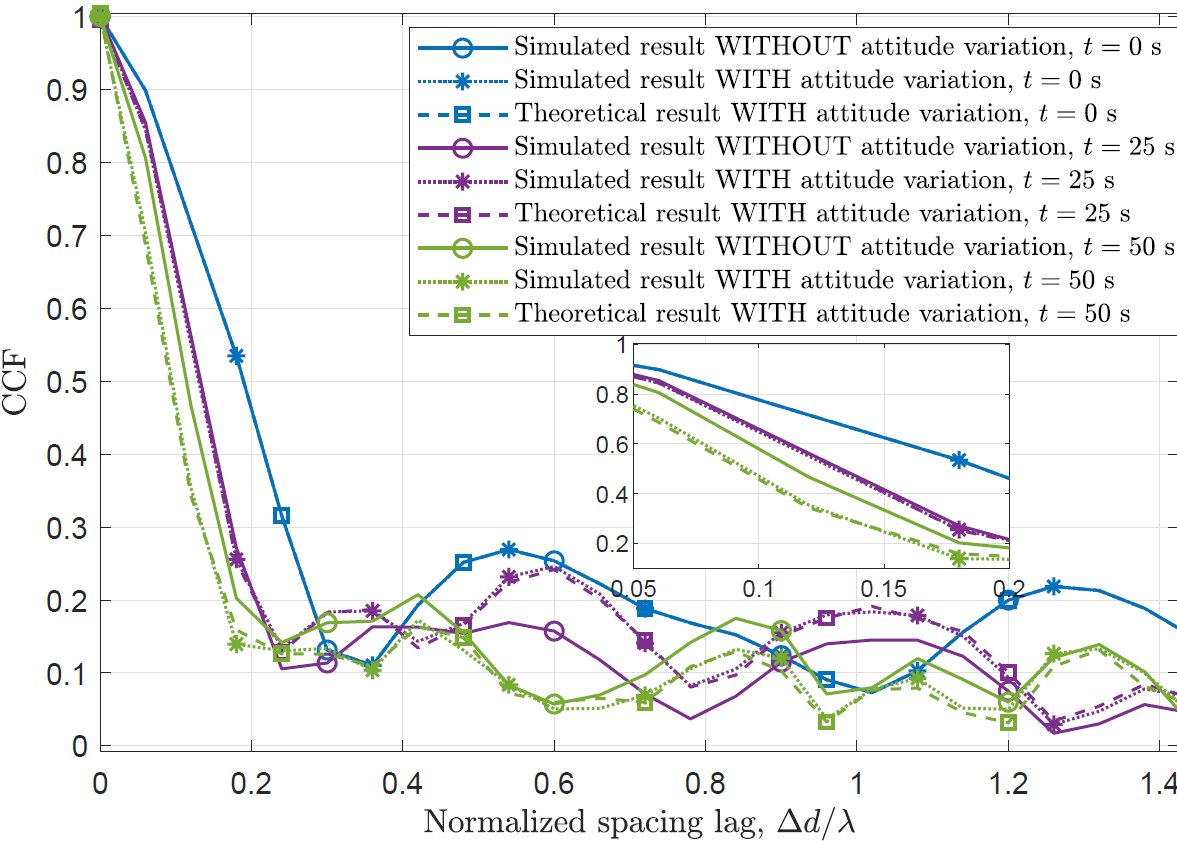}
	\caption{The time-variant CCFs with/without attitude variation (urban scenario,${f_0}$ = 28 GHz, $\left\| {{{\bf{v}}^{{\rm{tx}}}}} \right\|$ = 30 m/s, $\left\| {{{\bf{v}}^{{\rm{rx}}}}} \right\|$ = 5 m/s, $\gamma {|_{t = 0s}} = {0^ \circ }$, $\gamma {|_{t = 25s}} = {45^ \circ }$, $\gamma {|_{t = 50s}} = {90^ \circ }$).}
    \label{fig:11}
\end{figure}
\par To further verify the consistency of proposed model and generation method, it is necessary to compare the generated result with field-measurement data. To the best of our knowledge, so far only the authors in \cite{Guank20_Access} analyzed the ACF based on the measurement data for UAV mmWave communications. For the measurement scenario, the RX is placed on a building roof with the height of 25 m and the UAV performs five round-trip flights with the height of 0$\sim$24 m at the constant velocity of 1 m/s. The propagation link is the NLoS case when the UAV is below 11 m, and becomes the LoS case when it is over 11 m. The measurement campaigns are carried out at 1 GHz, 4 GHz, 12 GHz, and 24 GHz, respectively. In this paper, we mainly use the measured data at 24 GHz for validation. Moreover, the investigated ACF mainly focused on the relationship with UAV heights, thus (34) can be simplified to a distance-relative form. By setting ${f_0}$ = 24 GHz, $\left\| {{{\bf{v}}^{{\rm{tx}}}}} \right\|$ = 1 m/s, $\left\| {{{\bf{v}}^{{\rm{rx}}}}} \right\|$ = 0 m/s, and ${{\bf{R}}^{\rm{P}}}$ as identity matrix, the UAV-to-ground channel under the similar scenario \cite{Guank20_Access} can be generated based on the proposed model. The simulated ACF is compared with the measured one in Fig. 13, which achieves a good agreement with each other.

\begin{figure}[!b]
	\centering
	\includegraphics[width=88mm]{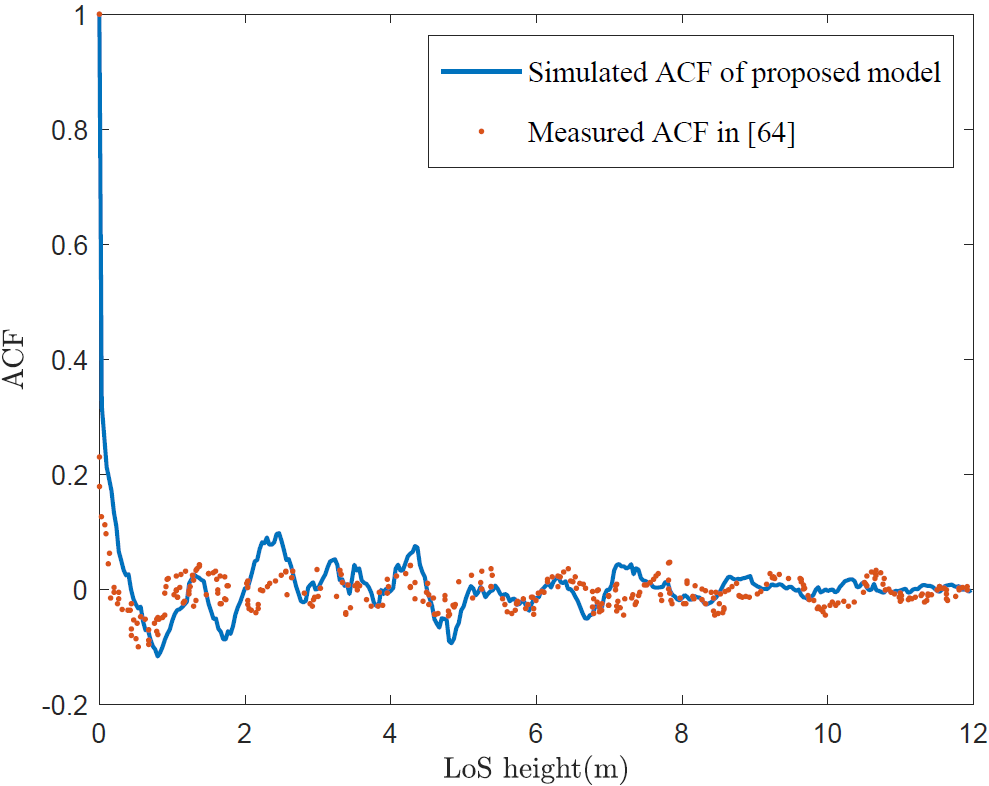}
	\caption{The simulated and measured ACFs (semi-urban scenario, ${f_0}$ = 24 GHz, $\left\| {{{\bf{v}}^{{\rm{tx}}}}} \right\|$ = 1 m/s, $\left\| {{{\bf{v}}^{{\rm{rx}}}}} \right\|$ = 0 m/s.}
    \label{fig:12}
\end{figure}

\section{Conclusions}
This paper has proposed a ML-based 3D non-stationary U2V mmWave channel model which takes into account of the velocity variation of  transceivers and attitude variation of UAV. Meanwhile, several ML-based networks, i.e., BPNN, K-means, and GAN are developed to train and generate the channel parameters, which can guarantee the consistence with geometric environment and obtain a good balance between accuracy and efficiency. The expressions of statistical properties for the proposed model, i.e., PDP, ACF, DPSD, and CCF have also been derived. Finally, the proposed method has been applied to generate the U2V mmWave channel under a typical urban scenario at 28 GHz. The generated PDP and DPSD have shown a good agreement with the RT-based result and the generated CCF and ACF have been compared with the theoretical and measurement results, which validates the effect of proposed method. Moreover, the results have demonstrated that the 3D rotation of UAV has an impact on CCF. In the future, we will carry out more field measurements for U2V mmWave communications and evaluate the communication system performance with controllable attitude variation of UAV.


\ifCLASSOPTIONcaptionsoff
  \newpage
\fi

\vspace{-4em}
\begin{IEEEbiography}[{\includegraphics[width=1in,height=1.25in,clip,keepaspectratio]{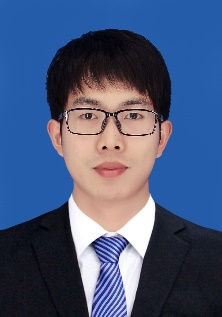}}]{Kai Mao} received the B.S. in information engineering and M.S. degree in electronics and communication engineering from the from the Nanjing University of Aeronautics and Astronautics (NUAA), Nanjing, China, in 2016 and 2019, respectively. He is currently pursuing the Ph.D. degree in communication and information systems. His research interests include channel sounding, modeling for UAV communication systems and wireless channel emulators.
\end{IEEEbiography}
\vspace{-4em}
\begin{IEEEbiography}[{\includegraphics[width=1in,height=1.25in,clip,keepaspectratio]{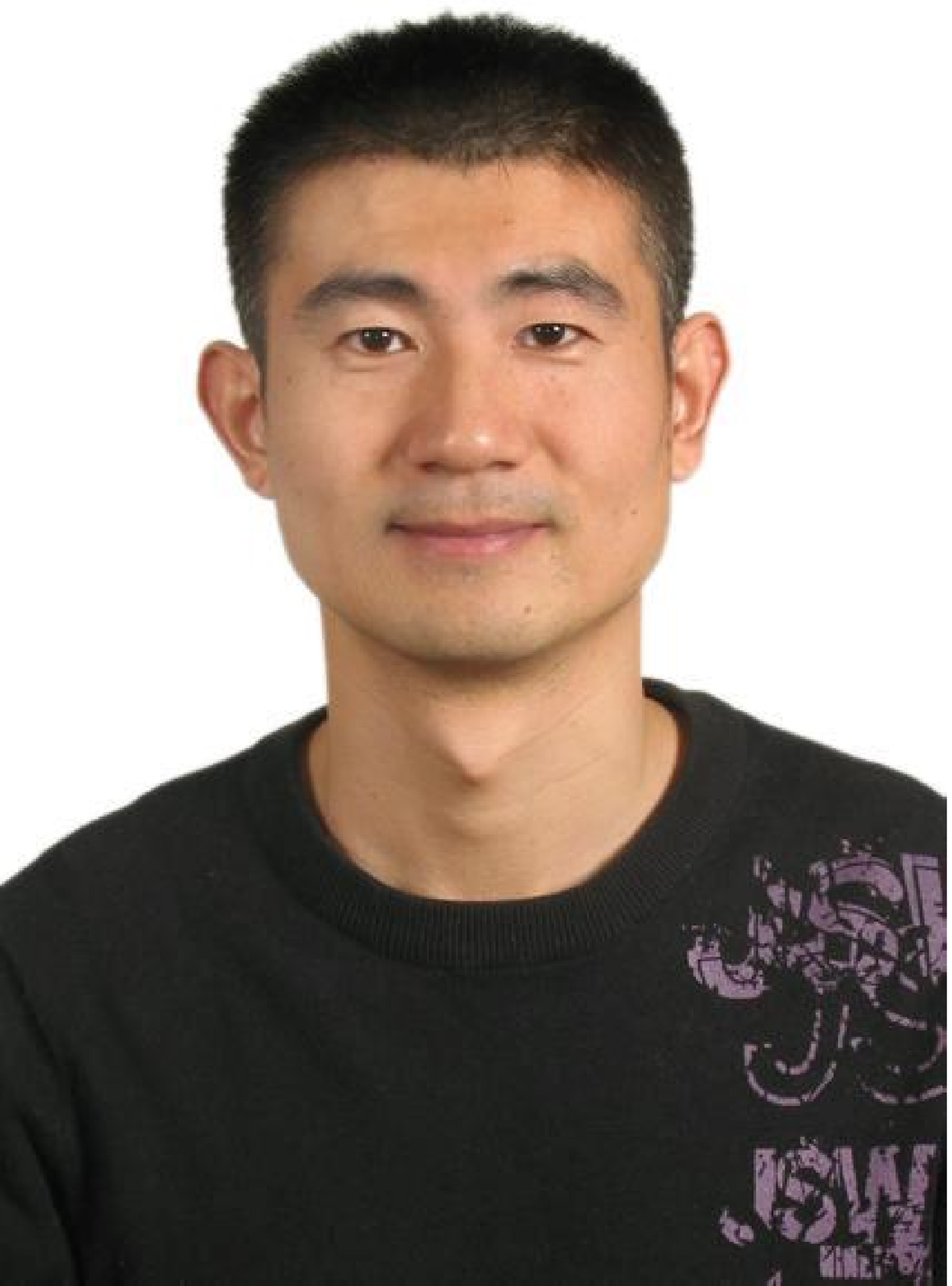}}]{Qiuming Zhu} received his B.S. degree in electronic engineering from Nanjing University of Aeronautics and Astronautics (NUAA) in Nanjing, China in 2002 and his M.S. and Ph.D. degrees in communication and information systems in 2005 and 2012, respectively. Since 2012, he has been an associate professor in wireless communications. From 2016 to 2017, he was also a visiting academic at Heriot-Watt University. His research interests include channel modeling for 5G communication systems and wireless channel emulators.
\end{IEEEbiography}
\vspace{-4em}
\begin{IEEEbiography}[{\includegraphics[width=1in,height=1.25in,clip,keepaspectratio]{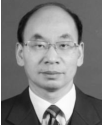}}]{Maozhong Song} received the master's degree in communications and electronic system from Zhejiang University, China, in 1986. He has been with the College of Electronic and Information Engineering, Nanjing University of Aeronautics and Astronautics,since1986,where he is currently a Professor. His research interests include wireless communications and satellite navigation, with a focus on modulation signal design, signal process-ing, signal simulator, and embedded systems and its applications in the Internet of Things.
\end{IEEEbiography}
\vspace{-4em}
\begin{IEEEbiography}[{\includegraphics[width=1in,height=1.25in,clip,keepaspectratio]{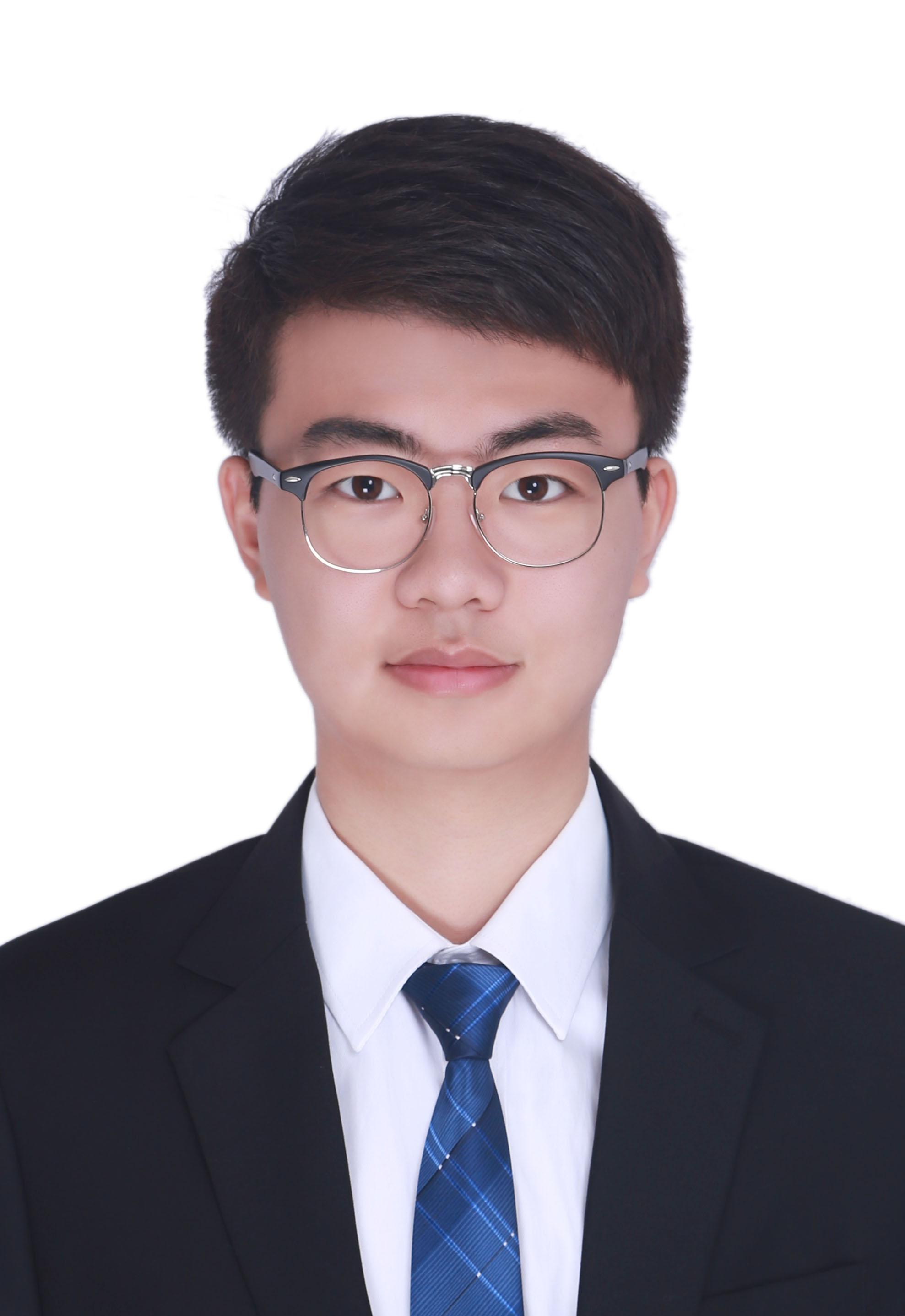}}]{Hanpeng Li} received his B.S. degree in software engineering from Northeastern University (NEU), Shenyang, China, in 2019.He is currently pursuing the M.S. degree in electrical and communications engineering at Nanjing University of Aeronautics and Astronautics (NUAA), Nanjing, China. His research interests include machine learning, deep learning, channel sounding and parameter estimation.
\end{IEEEbiography}
\vspace{10em}
\begin{IEEEbiography}[{\includegraphics[width=1in,height=1.25in,clip,keepaspectratio]{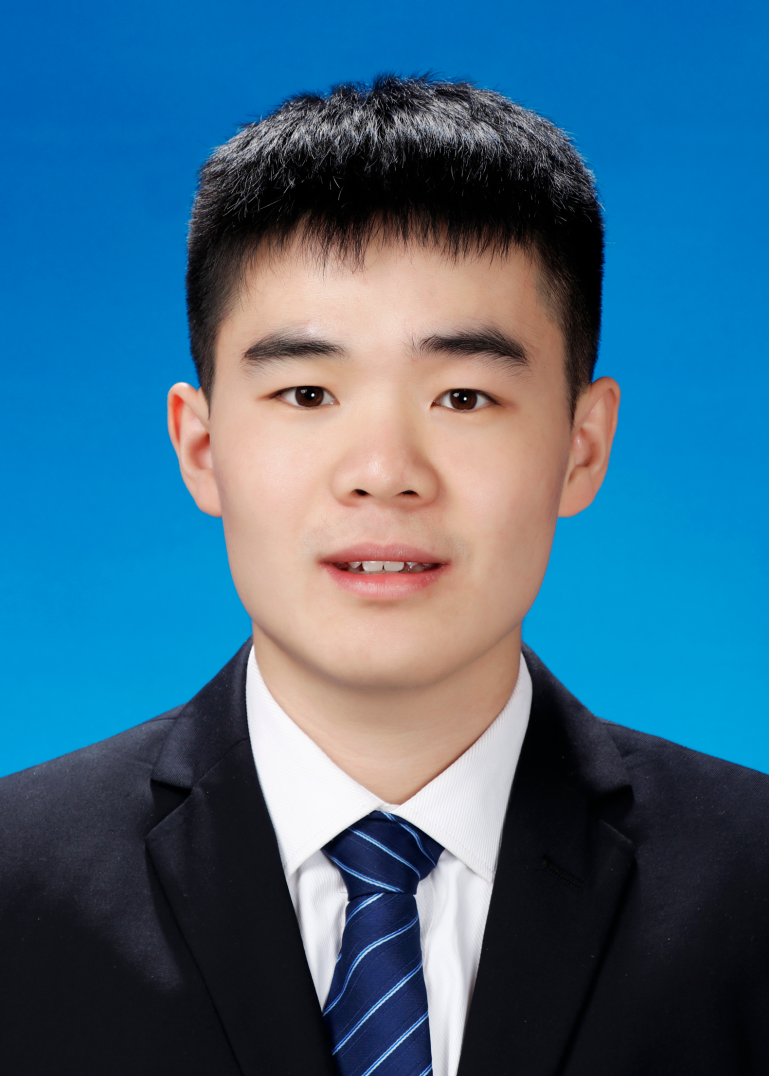}}]{Benzhe Ning} received his B.S. degree in communications engineering from North University of China (NUC), Taiyuan, China, in 2019.He is currently pursuing the M.S. degree in electrical and communications engineering at Nanjing University of Aeronautics and Astronautics (NUAA), Nanjing, China. His research interests include channel sounding, UAV communication system design and parameter estimation algorithm research.
\end{IEEEbiography}
\vspace{-35em}
\begin{IEEEbiography}[{\includegraphics[width=1in,height=1.25in,clip,keepaspectratio]{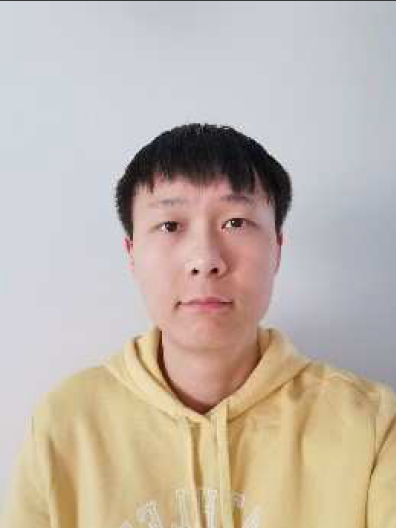}}]{Boyu Hua} received the B.S. degree in physics from Nanjing Normal University, China, in 2014, and the M.S. degree in electronic communication engineering from Nanjing University of Aeronautics and Astronautics (NUAA) , China, in 2018. Since 2018, he has been an experimentalist and is currently pursuing Ph.D. degree in communication and information systems in NUAA. His research interests include wireless channel modeling for 5G and B5G communication.
\end{IEEEbiography}
\vspace{-35em}
\begin{IEEEbiography}[{\includegraphics[width=1in,height=1.25in,clip,keepaspectratio]{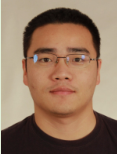}}]{Wei Fan} received his Bachelor of Engineering degree from Harbin Institute of technology, China in 2009, Master's double degree with highest honours from Politecnico di Torino, Italy and Grenoble Institute of Technology, France in 2011, and Ph.D. degree from Aalborg University, Denmark in 2014. From February 2011 to August 2011, he was with Intel Mobile Communications, Denmark as a research intern. He conducted a three-month internship at Anite telecoms oy, Finland in 2014. He is currently an associate professor at the Antennas, Propagation and Millimeter-wave Systems (APMS) Section at Aalborg University. His main areas of research are over-the-air testing of multiple antenna systems, radio channel sounding, modeling and emulation.
\end{IEEEbiography}
%








\end{document}